\documentstyle[12pt, hyperref]{article}

\catcode`\@=11
\def\theequation{\thesubsection.\arabic{equation}}
\@addtoreset{footnote}{section}
\@addtoreset{footnote}{subsection}
\@addtoreset{equation}{subsection}
\catcode`\@=12
\catcode`\@=11

\newcounter{app}

\def\app{\setcounter{equation}{0}
\def\theequation{A\arabic{app}.\arabic{equation}}\par
   \addvspace{4ex}
   \@afterindentfalse
  \secdef\@app\@dapp}
\newcommand\@app{\@startsection {app}{1}{0ex}%
                                   {-3.5ex \@plus -1ex \@minus -.2ex}%
                                   {2.3ex \@plus.2ex}%
                                   {\normalfont\Large\bf}}

\def\@dapp#1{%
{\parindent \z@ \raggedright  \bf #1}\par\nobreak}
\def\l@app#1#2{\ifnum \c@tocdepth >\z@
    \addpenalty\@secpenalty
    \addvspace{1.0em \@plus\p@}%
    \setlength\@tempdima{8.5em}%
    \begingroup
      \parindent \z@ \rightskip \@pnumwidth

      \parfillskip -\@pnumwidth
      \leavevmode \bfseries
      \advance\leftskip\@tempdima
      \hskip -\leftskip
      #1\nobreak\hfil \nobreak\hb@xt@\@pnumwidth{\hss #2}\par
    \endgroup\fi}
\catcode`\@=12

\small  \oddsidemargin -10mm  \evensidemargin 5mm  \topmargin -10mm
\def\be{\begin{equation}}
\def\ee{\end{equation}}

\renewcommand{\l}{\langle}
\renewcommand{\r}{\rangle}

\newtheorem{prop}{Proposition}
\newtheorem{theorem}{Theorem}
\newtheorem{remark}{Remark}
\newtheorem{Lemma}{Lemma}

\def\bprop{\begin{prop}}
\def\eprop{\end{prop}}
\def\bremark{\begin{remark}}
\def\eremark{\end{remark}}
\begin{document}
\author{A. Yu. Orlov\thanks{Oceanology Institute, Nahimovskii
prospect 36, Moscow, Russia, email address:
orlovs@wave.sio.rssi.ru}}
\title{Hypergeometric tau functions $\tau({\bf t},T,{\bf t}^*)$ as
$\infty$-soliton tau function in $T$ variables}
\date{}
\maketitle

\begin{abstract}
We consider KP tau function of hypergeometric type $\tau({\bf
t},T,{\bf t}^*)$, where the set ${\bf t}$ is the KP higher times
and $T,{\bf t}^*$ are sets of parameters. Fixing ${\bf t}^*$, we
find that $\tau({\bf t},T,{\bf t}^*)$ is an infinite-soliton
solution of different (dual) multi-component KP (and TL)
hierarchy, where the roles of the variables ${\bf t}$ and $T$ are
interchanged. When $\tau({\bf t},T,{\bf t}^*)$ is a polynomial in
${\bf t}$, we obtain a $N$-soliton solution of the dual hierarchy.
Parameters of the solitons are related to the Frobenius
coordinates of partitions in the Schur function development of
$\tau({\bf t},T,{\bf t}^*)$.
\end{abstract}
\tableofcontents

\section{Introduction}
 Hypergeometric tau functions \cite{nl64},\cite{pd22},\cite{pa24} were
introduced to generalize a notion of hypergeometric functions of
matrix argument \cite{GR},\cite{V}. These tau functions are series
over partitions of the form
\begin{equation}\label{htf}
\tau(n,{\bf t},T,{\bf t}^*)=c_n(T)\sum_{\lambda \in P}
e^{\sum_{i=1}^\infty T_{n-i}-T_{n+\lambda_i-i}}s_\lambda({\bf
t})s_\lambda({\bf t}^*) \ ,
\end{equation}
where  $T=\{T_m,m=0,\pm 1,\pm 2,\dots \}$ are arbitrary
parameters, the $s_\lambda$ denotes the Schur function, and $P$ is
the set of all partitions. The factor $c_n(T)=\tau (n,{\bf
0},{T},{\bf 0} )^{-1}$ is not important in the context of the
present paper and is given by (\ref{H0+})-(\ref{H0-}) in the
Appendix.

Considered as a function of the variables ${\bf
t}=(t_1,t_2,\dots)$, series (\ref{htf}) is a KP tau function,
where the set ${\bf t}$ plays the role of KP higher times; the
second set of parameters ${\bf t}^*$ is related to a second KP
hierarchy. As a function of $n,{\bf t},{\bf t}^*$, series
(\ref{htf}) is a Toda lattice (TL) tau function. If the variables
$T$ solve the periodicity condition $T_i=T_{i+N}$, then
(\ref{htf}) is a tau function of the so-called N-periodic TL.
Certain known hypergeometric functions of many variables
(considered in \cite{GR},\cite{V}) can be obtained by specializing
the variables ${\bf t},T$, and ${\bf t}^*$. We note that the
hypergeometric tau functions yields a perturbative asymptotic
expansion for a set of known matrix integrals
\cite{1},\cite{Or},\cite{Cadiz}. They were also used to construct
 new solvable matrix integrals \cite{1'}.

A special interest in this kind of tau functions is in a set of
applications. Tau functions arising in supersymmetric gauge
theories \cite{N},\cite{LMN}, in the problem of counting of
Hurwitz numbers \cite{O},  of counting of Gromov-Witten invariants
of $P^1$ \cite{OP}, of computation of intersection numbers on
Hilbert schemes \cite{LQW} - are of the hypergeometric type. Tau
functions, which were considered in \cite{NTT},\cite{T} in the
context of $c=1$ strings, are also of the hypergeometric type.

In the present paper, we specialize the variables ${\bf t}^*$, and
study  tau function (\ref{htf}) as a function of the variables
$T$. We succeed to do it when ${\bf t}^*$ belongs to any of four
special families, given by
(\ref{choicetinftyp}),(\ref{choicet(a)p}), (\ref{choicetinftyqp}),
and (\ref{choicet(a)qp}) below. We find that series (\ref{htf}) is
a multi-soliton tau function of a different integrable hierarchy,
which we call a dual hierarchy (the type of the hierarchy is
defined by the choice of ${\bf t}^*$). The variables $T $ of
(\ref{htf}) turn out to be linear combinations of the time
variables ${\tilde n}^{(m)},{\tilde {\bf t}}^{(m)},{\tilde {\bf
t}}^{(m)*},m=1,\dots,p$ of the new hierarchy (namely, it is the
so-called $p$ component TL (and KP) hierarchy). Sometimes, we omit
the arguments $n$ and ${\bf t}^*$ and use the notations $\tau({\bf
t},T)$ and $\tau({\bf t},T,{\bf t}^*)$, instead of $\tau(n,{\bf
t},T,{\bf t}^*)$.

The present letter is a development of the preprint \cite{hypsol}.

First, we briefly review some facts and notions.

{\bf Soliton theory}. The KP hierarchy of integrable equations
\cite{ZSh},\cite{JM}, which is the most popular example in the
soliton theory, consists of the semi-infinite set of nonlinear
integral-differential evolutionary equations
\begin{equation}\label{KPh}
\partial_{t_m}u=K_m[u] \ ,\quad m=1,2,\dots \ ,
\end{equation}
which are commuting flows: $
\left[\partial_{t_k},\partial_{t_m}\right]u=0 $. The first
nontrivial one is the Kadomtsev-Petviashvili equation
\begin{equation}\label{KP}
\partial_{t_3}u=\frac 14 \partial_{t_1}^3 u+ \frac 34
\partial_{t_1}^{-1}\partial_{t_2}^2u+\frac 34 \partial_{t_1} u^2 \ ,
\end{equation}
which, originally, served in the plasma physics \cite{ZSh}, now
plays a very important role both, in physics  and in mathematics.
Another very important equation is the equation of two-dimensional
Toda lattice (TL)
\begin{equation}\label{Toda'}
\partial_{t_1}\partial_{t^*_1}\phi_n=
e^{\phi_{n-1}-\phi_{n}}- e^{\phi_{n}-\phi_{n+1}}
\end{equation}
first integrated in \cite{AMi} and carefully studied in \cite{UT}
in the framework of \cite{JM}. This equation gives rise to TL
hierarchy which contains derivatives with respect to the higher
times $t_1,t_2,\dots $ and $t_1^*,t_2^*,\dots \ $.

The key point of the soliton theory, at present, is a notion of
tau function, introduced by Sato (for the KP tau-function see
\cite{JM}). The tau function is a sort of a potential, which
respectively gives rise both to the TL and KP hierarchies. It
depends on two semi-infinite sets of the higher times
$t_1,t_2,\dots$ and $t_1^*,t_2^*,\dots$, and on the discrete
variable $n$: $\tau=\tau(n,{\bf t},{\bf t^*})$. More explicitly,
we have \cite{JM},\cite{UT}:
\begin{equation}\label{tauuphi}
\quad u=2\partial_{t_1}^2\log \tau(n,{\bf t},{\bf t^*}) \ ,\quad
\phi_n({\bf t},{\bf t}^*)=-\log \frac{\tau(n+1,{\bf t},{\bf
t}^*)}{\tau(n,{\bf t},{\bf t}^*)}
\end{equation}

Three examples of tau functions.

(1) {\bf The vacuum TL tau function} is
\begin{equation}\label{vactau}
\tau({\bf t},{\bf t}^*)=e^{\sum_{m=1}^\infty mt_mt_m^*}
\end{equation}
This is the simplest example of (\ref{htf}), where $T=0$.

(2) {The \bf N-soliton tau function} of the TL and the KP
hierarchies are defined by a set of parameters $\{p_i,q_i,a_i\}\
,i=1,\dots,N ,\ p_i\neq q_i$; the parameters $p_i,q_i$ determines
the velocity of the soliton labeled by a number $i$; while the
$a_{i}$ determines the initial location of the soliton number $i$.
The N-soliton KP and TL tau function is the series
\begin{equation}\label{KPsoliton}
\tau^{solitons}(n,{\bf t},{\bf t}^*) =c({\bf t},{\bf
t}^*)\left(\sum_{p=0}^N \ \sum_{1\le i_1<\cdots <i_p}\
\prod_{k<m}c_{i_ki_m}e^{\eta_{i_1}+\cdots +\eta_{i_p}}\right)=
\end{equation}
\begin{equation}
c({\bf t},{\bf t}^*)\left(1+\sum_{i=1}^N e^{\eta_i}+\sum_{1\le
i<j}^N c_{ij}e^{\eta_i+\eta_j}+\cdots\right),
\end{equation}
where
\begin{equation}\label{eta}
\eta_i=\xi(n,{\bf t},{\bf t}^*,p_i)-\xi(n,{\bf t},{\bf
t}^*,q_i)+\log \frac{a_{i}}{p_i-q_i}\ ,
\end{equation}
\begin{equation}\label{xi}
\xi(n,{\bf t},{\bf t}^*,p)=n\log p+\sum_{m=1}^\infty t_m
p^m-\sum_{m=1}^\infty t^*_m p^{-m}\ ,
\end{equation}
\begin{equation}\label{cij}
c_{ij}=\frac{(p_i-p_j)(q_i-q_j)}{(p_i-q_j)(q_i-p_j)}
\end{equation}
and
\begin{equation}\label{ctt*}
c({\bf t},{\bf t}^*)=e^{\sum_{m=1}^\infty mt_mt_m^*}
\end{equation}
is the vacuum TL tau function. The series inside the brackets in
(\ref{KPsoliton}) terminates when the number of solitons, $N$, is
finite. If we are interested in the KP case, the values of the
variables ${\bf t}^*$ are irrelevant.

We are also interested in the fermionic representation of this
soliton solution \cite{JM} (see Appendix A for notations):
\begin{equation}\label{KPsolitonferm}
\tau^{solitons}(n,{\bf t},{\bf t}^*)= \l n|e^{H({\bf t})}
e^{\sum_{i=1}^N a_{i}\psi(p_i)\psi^*(q_i)}e^{H^*({\bf t}^*)} |n\r
\end{equation}
In what follows, we are interested in a degenerate case of soliton
tau function (\ref{KPsolitonferm}) ('resonances'), the case when
some of the $p_i$ are identical (and the same about the variables
$q_i$), as described in (\ref{inftysol}) below.

(3) {\bf The polynomial TL tau functions}, which yields rational
solutions of the TL hierarchy, were found by K.Takasaki. He found
rather general class of solutions of the TL hierarchy in the form
of double series in the Schur functions over partitions
(definitions see below) \cite{Tinit}:
\begin{equation}\label{tauschurschur}\label{Takasaki}
\tau(n,{\bf t},{\bf t^*})=\sum_{\lambda,\mu}K_{\lambda
\mu}(n)s_\lambda({\bf t})s_\mu ({\bf t^*}) \ ,
\end{equation}
where the coefficients $K_{\lambda \mu }$  can be presented as
certain determinants or, alternatively, just solve special
bilinear equations \cite{JM}. Series (\ref{Takasaki}) gives rise
to a general KP and TL solution, see \cite{Tinit}; the so-called
rational solutions one finds as special cases of (\ref{Takasaki}),
for instance, any polynomial tau function gives rise to a rational
KP and TL solution.

 We can show that series (\ref{htf}) is  a special example of
 Takasaki series (\ref{tauschurschur}). Our goal is to show that
series (\ref{htf}) (in particular, the terminating ones, which
gives rise to a rational solution of the KP hierarchy) can be
viewed as a special soliton solution of a different ("dual")
hierarchy in the case when we fix the parameters ${\bf t}^*$ in
one of the ways described below by
(\ref{choicetinfty}),(\ref{choicet(a)}),(\ref{choicetinftyq}),
(\ref{choicet(a)q}) and by
(\ref{choicetinftyp}),(\ref{choicet(a)p}),(\ref{choicetinftyqp}),
(\ref{choicet(a)qp}). These are exactly the specializations of the
parameters ${\bf t}^*$, which we find in many applications of
hypergeometric type series (\ref{htf}).
   By now, we failed to
generalize our results for a different specialization of ${\bf
t}^*$, and for the Takasaki series  of general type
(\ref{Takasaki}).

{\bf Partitions}. Polynomial functions of many variables are
parameterized by partitions. A {\em partition} is any (finite or
infinite) sequence of non-negative integers in the decreasing
order:
\begin{equation}\label{partition}
\lambda = (\lambda_1, \lambda_2, \dots,\lambda_r,\dots )\ ,\quad
\lambda_1 \ge \lambda_2 \ge \dots \ge\lambda_r \ge \dots \ge 0
\end{equation}
The numbers $\lambda_i$ in (\ref{partition}) are called the {\em
parts} of the $\lambda$. The number of the non-vanishing parts is
the {\em length} of the $\lambda$, denoted by $\ell(\lambda)$. The
sum of the parts is the {\em weight} of the $\lambda$, denoted by
$|\lambda|$. If $n=|\lambda|$, we say that the $\lambda$ is a {\em
partition of} $n$. The zero partition (for which $\lambda_1=0$) is
denoted by $0$.

The {\em diagram} of a partition (or the {\em Young diagram}) may
be defined as the set of points (or nodes) $(i,j) \in Z^2$ such
that $1\le j \le \lambda_i$. Thus, the Young diagram is viewed as
a subset of entries in a matrix with the $l(\lambda)$ lines and
the $\lambda_1$ rows. We denote the diagram of $\lambda$ by the
same symbol $\lambda$.

For example,

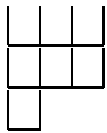
\vbox{

\qquad\qquad\qquad\qquad\qquad\qquad\qquad\begin{tabular}{|c|}
   \\ \hline
\end{tabular}\begin{tabular}{|c|}
   \\ \hline
\end{tabular}\begin{tabular}{|c|}
   \\ \hline
\end{tabular}

\qquad\qquad\qquad\qquad\qquad\qquad\qquad\begin{tabular}{|c|}
   \\ \hline
\end{tabular}\begin{tabular}{|c|}
   \\ \hline
\end{tabular}\begin{tabular}{|c|}
   \\ \hline
\end{tabular}

\qquad\qquad\qquad\qquad\qquad\qquad\qquad\begin{tabular}{|c|}
   \\ \hline
\end{tabular}\\
\\
} is the diagram of $(3,3,1)$. The weight of this partition is
$7$, the length is equal to $3$.

The partition whose diagram is obtained by the transposition of
the diagram $\lambda$ with respect to the main diagonal is called
the conjugated partition and denoted by $\lambda'$.

Another notation is due to Frobenius. Suppose that the main
diagonal of the diagram of $\lambda$  consists of $r$ nodes
$(i,i)\quad (1\le i\le r)$. Let $\alpha_i=\lambda_i-i$ be the
number of nodes in the $i$th row of $\lambda$ to the right of
$(i,i)$, for $1\le i\le r$, and let $\beta_i=\lambda_i'-i$ be the
number of nodes in the $i$th column of $\lambda$ below $(i,i)$,
for $1\le i\le r$. We have $\alpha_1>\alpha_2>\cdots >\alpha_r\ge
0$ and $\beta_1>\beta_2>\cdots >\beta_r\ge 0$. Then, we denote the
partition $\lambda$ by
\begin{equation}\label{Frob}
\lambda = \left( \alpha_1,\dots ,\alpha_r|\beta_1,\dots
,\beta_r\right)=(\alpha |\beta )
\end{equation}
One may say that the Frobenius notation corresponds to a
decomposition of a diagram $\lambda$ into main hooks, where the
biggest hook is $\left( \alpha_1|\beta_1\right)$, next one is
$\left( \alpha_2|\beta_2\right)$, and so on up to the smallest
hook, which is $\left( \alpha_r|\beta_r\right)$. The corners of
the main hooks are situated at the main diagonal of the diagram.
For instance, the
partition $(3,3,1)$ consists of the two main hooks $(2,2)$ and $(1,0)$:\\

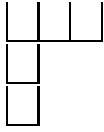
\vbox{

\qquad\qquad\begin{tabular}{|c|}
   \\ \hline
\end{tabular}\begin{tabular}{|c|}
   \\ \hline
\end{tabular}\begin{tabular}{|c|}
   \\ \hline
\end{tabular}

\qquad\qquad\begin{tabular}{|c|}
   \\ \hline
\end{tabular}
\qquad\qquad\qquad\qquad and
\qquad\qquad\qquad\qquad\begin{tabular}{|c|}
   \\ \hline
\end{tabular}\begin{tabular}{|c|}
   \\ \hline
\end{tabular}

\qquad\qquad\begin{tabular}{|c|}
   \\ \hline
\end{tabular}
\\
\\
}

 In the Frobenius notation this is $(2,1|2,0)$. If
 $\lambda=(\alpha|\beta)$, then $\lambda'=(\beta|\alpha)$

{\bf The $\tau$ functions of hypergeometric type}.

 Let us consider a function $r$,
which depends on a single variable $n$, the $n$ is an integer.
Given partition $\lambda$, we define
\begin{equation}\label{rlambda}
r_\lambda(x)=\prod_{i,j\in \lambda}r(x+j-i)
\end{equation}
Namely, the $r_\lambda(n)$ is the product of the $r$ over all
nodes of Young diagram of the partition $\lambda$, at each node,
the argument of the $r$ is defined by the entry, $i,j$, of the
node. The number $j-i$ is zero on the main diagonal; the $j-i$ is
called the content of the node with an entry  $i,j$.

 For instance, for the partition $(3,3,1)$ the diagram is

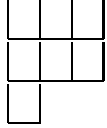
\vbox{

\qquad\qquad\qquad\qquad\qquad\qquad\qquad\begin{tabular}{|c|}
   \\ \hline
\end{tabular}\begin{tabular}{|c|}
   \\ \hline
\end{tabular}\begin{tabular}{|c|}
   \\ \hline
\end{tabular}

\qquad\qquad\qquad\qquad\qquad\qquad\qquad\begin{tabular}{|c|}
   \\ \hline
\end{tabular}\begin{tabular}{|c|}
   \\ \hline
\end{tabular}\begin{tabular}{|c|}
   \\ \hline
\end{tabular}

\qquad\qquad\qquad\qquad\qquad\qquad\qquad\begin{tabular}{|c|}
   \\ \hline
\end{tabular}\\
\\
}

so, our $r_\lambda(x)$ is equal to
$r(x+2)(r(x+1))^2(r(x))^2r(x-1)r(x-2)$.

For zero partition we set $r_0 \equiv 1$.

It was shown \cite{tmf} that
\begin{equation}\label{exex'}
\tau_r(n,{\bf t},{\bf t}^*)=\sum_\lambda r_\lambda (n)
s_\lambda({\bf t})s_\lambda({\bf t^*})
\end{equation}
(where the sum ranges all partitions including the zero partition)
is a TL and a KP tau function, which we call the tau function of
hypergeometric type. Via the change of variables
\begin{equation}\label{rT}
 r(m)=e^{T_{m-1}-T_{m}}
\end{equation}
we rewrite it in form (\ref{htf}).

The Schur functions $s_\lambda({\bf t}),s_\lambda({\bf t^*})$ are
defined with the help of
\begin{equation}\label{Schurtt*}
s_\lambda({\bf t})=\det h_{\lambda_i-i+j}({\bf t})_{1\le i,j\le
l(\lambda) } \ ,\quad \exp\sum_{m=1}^\infty
z^mt_m=\sum_{k=0}^\infty z^kh_k({\bf t}) \ ,
\end{equation}
and for $k<0$, we set $h_k=0$ . Here, the $h_k({\bf t})$ is called
the elementary Schur function, or, the same, the complete
symmetric function,  see \cite{Mac}.

For instance, if we choose the $r$ to be rational function, then
for ${\bf t}^*=(1,0,0,\dots ) $ and for $ t_m=tr \ {\bf X}^m\,\
m=1,2,\dots $, we verify  \cite{nl64},\cite{pd22}, that we obtain
a known hypergeometric function, which is called the
hypergeometric function of matrix argument, see (\ref{hZ1}).

The equation
\begin{equation}\label{rToda'}
\partial_{t_1}\partial_{t^*_1}\phi_n=
r(n)e^{\phi_{n-1}-\phi_{n}}- r(n+1)e^{\phi_{n}-\phi_{n+1}} \ ,
\end{equation}
(which is similar to the Toda lattice equation) holds for
\begin{equation}\label{phi_n'}
\phi_n({\bf t},{\bf t}^*)=-\log \frac{\tau_r(n+1,{\bf t},{\bf
t}^*)}{\tau_r(n,{\bf t},{\bf t}^*)}
\end{equation}

The tau function $\tau_r(n,{\bf t},{\bf t}^*)$ and the tau
function $\tau(n,{\bf t},{T},{\bf t}^*)$, see Appendix A, are
solutions of (\ref{rToda'}) and of (\ref{Toda}) respectively,
these tau functions are identical up to the multiplication by the
number $c_n$, independent of ${\bf t},{\bf t}^*$, see
(\ref{tauhyp'}). So, it is enough to study one of them, say,
$\tau(n,{\bf t},T,{\bf t}^*)$.

\section{The tau-functions $\tau({\bf t},T)$ as a soliton
tau function of a dual hierarchy}

First of all, we rewrite  sum (\ref{htf}), using the Frobenius
notations $(\alpha|\beta)$ for each partition $\lambda$;
$(\alpha|\beta)=(\alpha_1,\dots ,\alpha_k|\beta_1,\dots,\beta_k)
$, where $k$ is the number of the main hooks in the diagram of a
partition $\lambda$. We restate (\ref{htf}) as
\begin{equation}\label{KPsoliton'}
\tau(n,{\bf t},T,{\bf t}^*)=c_n(T)\left(
1+\sum_{k=1}^\infty\sum_{\alpha_1>\cdots>\alpha_k\ge0 \atop
\beta_1>\cdots>\beta_k\ge0 }^\infty e^{\sum_{i=1}^k
(T_{n-\beta_i-1}-T_{n+\alpha_i})}s_{(\alpha|\beta)}({\bf
t})s_{(\alpha|\beta)}({\bf t}^*)\right)
\end{equation}
\begin{remark}. Actually, one can restrict the sum over the infinite
sets of the non-negative integers $\{\alpha_i \in {Z}_{\ge 0},\
i=1,2,\dots \}$ (and $\{\beta_i \in {Z}_{\ge 0},\ i=1,2,\dots$\})
to any subset $S_\alpha \subseteq {Z}_{\ge 0}$ (respectively,
$S_\beta \subseteq {Z}_\ge$). It means, that the sum range not all
partitions, but only the set we denote by $S$,
\begin{equation}\label{subseth}
\tau(n,{\bf t},T,{\bf t}^*)=c_n(T) \sum_{\lambda \in S}
e^{\sum_{i=1}^\infty T_{n-i}-T_{n+\lambda_i-i}}s_\lambda({\bf
t})s_\lambda({\bf t}^*)
\end{equation}
\begin{equation}\label{subsetFrob}
=c_n(T)\left(1+\sum_{k=1}^\infty\sum_{\alpha_1>\cdots>\alpha_k\ge0
\atop \alpha_i \in S_\alpha }\sum_{\beta_1>\cdots>\beta_k\ge0
\atop \beta_i \in S_\beta }^\infty e^{\sum_{i=1}^k
(T_{n-\beta_i-1}-T_{n+\alpha_i})}s_{(\alpha|\beta)}({\bf
t})s_{(\alpha|\beta)}({\bf t}^*)\right)
\end{equation}
In the case when $S\neq P$, for simplicity we set
 $n=0$; the detailed study see in the forthcoming
paper.
\end{remark}

We introduce the notations:

\begin{equation}\label{choicetinfty}
{\bf t}_\infty=(1,0,0,0,\dots) \ ,
\end{equation}
\begin{equation}\label{choicet(a)}
{\bf t}(a)=(\frac{a}{1},\frac{a}{2},\frac{a}{3},\dots) \ ,
\end{equation}
\begin{equation}\label{choicetinftyq}
{\bf t}(\infty,q)=(t_1(\infty,q),t_2(\infty,q), \dots ),\quad
t_m(\infty,q)= \frac{1}{m(1-q^m)} \ ,\quad m=1,2,\dots \ ,
\end{equation}
\begin{equation}\label{choicet(a)q}
{\bf t}(a,q)=(t_1(a,q),t_2(a,q),\dots ) \ ,\quad
t_m(a,q)=\frac{1-(q^a)^{m}}{m(1-q^m)}\ ,\quad m=1,2,\dots
\end{equation}

In the sequel, we fix ${\bf t}^*$  to be one of
(\ref{choicetinfty}),(\ref{choicet(a)}),(\ref{choicetinftyq}),
(\ref{choicet(a)q}).
\begin{Lemma}. Let $(\alpha|\beta)=(\alpha_1,\dots ,\alpha_k
|\beta_1,\dots ,\beta_k)$ be the Frobenius notation for a
partition. Then,
\begin{equation}\label{stinfty}
s_{(\alpha|\beta)}({\bf
t}_\infty)=\frac{\prod^k_{i<j}(\alpha_i-\alpha_j)(\beta_i-\beta_j)}
{\prod_{i,j=1}^k(\alpha_i+\beta_j+1)}\frac {1}{\prod_{i=1}^k
\alpha_i!\prod_{i=1}^k \beta_i!} \ ,
\end{equation}
\begin{equation}\label{st(a)}
s_{(\alpha|\beta)}({\bf
t}(a))=\frac{\prod^k_{i<j}(\alpha_i-\alpha_j)(\beta_i-\beta_j)}
{\prod_{i,j=1}^k(\alpha_i+\beta_j+1)} \prod_{i=1}^k\frac
{(a)_{\alpha_i+1}}{\alpha_i!} \prod_{i=1}^k
\frac{(-)^{\beta_i}(-a)_{\beta_i}}{ \beta_i!} \ ,
\end{equation}
\begin{equation}\label{stinftyq}
s_{(\alpha|\beta)}({\bf
t}(\infty,q))=\frac{\prod^k_{i<j}(q^{\alpha_i+1}-q^{\alpha_j+1})(q^{-\beta_j}-q^{-\beta_i})}
{\prod_{i,j=1}^k(q^{-\beta_i}-q^{\alpha_j+1})}\frac
{1}{\prod_{i=1}^k(q;q)_{\alpha_i}\prod_{i=1}^k (q;q)_{\beta_i}} \
,
\end{equation}
\begin{equation}
s_{(\alpha|\beta)}({\bf t}(a,q))=
\end{equation}
\begin{equation}\label{st(a,q)}
\frac{\prod^k_{i<j}(q^{\alpha_i+1}-q^{\alpha_j+1})(q^{-\beta_j}-q^{-\beta_i})}
{\prod_{i,j=1}^k(q^{-\beta_i}-q^{\alpha_j+1})} \prod_{i=1}^k\frac
{(q^a;q)_{\alpha_i +1}}{(q;q)_{\alpha_i}}\prod_{i=1}^k\frac
{(-)^{\beta_i}q^{(a-1)\beta_i }(q^{1-a};q)_{\beta_i}}
{(q;q)_{\beta_i}}
\end{equation}
\end{Lemma}

The proof follows from the formulae:
\begin{equation}\label{s=dets}
s_{(\alpha|\beta)}({\bf t})=\det s_{(\alpha_i|\beta_j)}({\bf
t})|_{i,j=1,\dots ,k}\ ,
\end{equation}
\begin{equation}
s_{(\alpha_i|\beta_j)}({\bf
t}_\infty)=\frac{1}{H_{(\alpha_i|\beta_j)}}=\frac{1}{(\alpha_i)!
(\beta_i)!(\alpha_i+\beta_j+1)}\ ,
\end{equation}
where the $H_{(\alpha_i|\beta_j)}$ is the product-of-hook-length
\cite{Mac} of the partition $(\alpha_i|\beta_j)$, and from
\begin{equation}\label{PochSchur'}
 (a)_{(\alpha|\beta)}=\frac{s_{(\alpha|\beta)}({\bf t}(a))}
{s_{(\alpha|\beta)}({\bf t }_\infty)} \ ,
\end{equation}
\begin{equation}\label{(a)lambda}
(a)_{(\alpha|\beta)}=\prod_{i=1}^k(-)^{\beta_i}(a)_{\alpha_i
+1}(1-a)_{\beta_i} \ ,
\end{equation}
where
\begin{equation}\label{pohh}
(a)_k\ :=\frac{\Gamma(a+k)}{ \Gamma(a)}=a(a+1)\cdots(a+k-1),
\end{equation}
and from the similar relations in the $q$-case:
\begin{equation}
s_{(\alpha_i|\beta_j)}({\bf t}(\infty,q))=\frac{q^{\frac 12
(\beta_i^2+\beta_i)}}{H_{(\alpha_i|\beta_j)}(q)}=\frac{q^{\frac 12
(\beta_i^2+\beta_i)}}
{(q;q)_{\alpha_i}(q;q)_{\beta_i}(1-q^{\alpha_i+\beta_j+1})}\ ,
\end{equation}
\begin{equation}\label{PochSchur'q}
(q^a;q)_{(\alpha|\beta)}=\prod_{(i,j)\in {(\alpha|\beta)}}
(1-q^{a+j-i})= \frac{s_{(\alpha|\beta)}({\bf
t}(a,q))}{s_{(\alpha|\beta)} ({\bf t }(\infty,q))} \ ,
\end{equation}
\begin{equation}\label{(a)lambdaq}
(q^a;q)_{(\alpha|\beta)}=\prod_{i=1}^k(-)^{\beta_i}q^{a\beta_i-\frac
12(\beta_i^2+\beta_i) }(q^a;q)_{\alpha_i +1}(q^{1-a};q)_{\beta_i}
\ ,
\end{equation}
where
\begin{equation}\label{pohhq}
(q^a;q)_k\ :=(1-q^a)(1-q^{a+1})\cdots (1-q^{a+k-1}) \ ,\quad
(q^a;q)_0\ :=1
\end{equation}

\quad

Next, let us consider a second TL hierarchy with the higher times,
which we denote by ${\tilde n},\tilde{{\bf t}},\tilde{{\bf t}}^*$.
We study {\bf the infinite number of  solitons tau function} of a
degenerate form
\begin{equation}\label{inftysol}
\tau^{solitons}({\tilde n},\tilde{{\bf t}},\tilde{{\bf t}}^*)=\l
{\tilde n}|e^{{H}(\tilde{{\bf t}})} e^{\sum_{i\ge0,j>0}^\infty
a_{ij}{\psi}(z_i){\psi^*}(z_{-j})} e^{{H^*}(\tilde{{\bf
t}}^*)}|{\tilde n}\r \ ,
\end{equation}
where all $z_i\ (i \in {\bf Z})$ are different, and
\begin{equation} \tilde{{\bf t}}=(\tilde{ t}_1,\tilde{ t}_2,\tilde{
t}_3,\dots ),\quad \tilde{{\bf t}}^*=(\tilde{ t}_1^*,\tilde{
t}_2^*,\tilde{ t}_3^*,\dots )
\end{equation}
(as we see, (\ref{inftysol}) is related to (\ref{KPsolitonferm}),
where there are identical numbers among the $p_i,i=1,2,\dots$, and
identical numbers among the $q_i,i=1,2,\dots$).

We develop (\ref{inftysol}) in the $a_{ij}$ variables and use
(\ref{xit}), (\ref{xit*}) and (\ref{fervan}). We obtain
\begin{equation}\label{deta}
\tau^{solitons}({\tilde n},\tilde{{\bf t}},\tilde{{\bf
t}}^*)e^{-\sum_{m=1}^\infty m{\tilde t}_m{\tilde t}^*_m}=
\end{equation}
\begin{equation}\label{deta'}
1+ \sum_{k=1}^\infty\sum_{{\alpha_1>\cdots>\alpha_k\ge0} \atop
{\beta_1> \cdots>\beta_k\ge 0}} \left(\det
a_{\alpha_i\beta_j}|_{i,j=1}^{k}\right) \l 0|
\prod_{i,j=1}^{k}e^{\xi(\tilde{n},\tilde{{\bf t}},\tilde{{\bf
t}}^*,z_{\alpha_i})-\xi(\tilde{n},{\tilde{\bf t}},{\tilde {\bf
t}}^*,z_{-\beta_j})} {\psi}(z_{\alpha_i}){\psi^*}(z_{-\beta_j})
|0\r
\end{equation}
\begin{equation}\label{deta''}
=1+\sum_{k=1}^\infty\sum_{{\alpha_1>\cdots>\alpha_k\ge0} \atop
{\beta_1> \cdots>\beta_k\ge 0}}e^{\xi(\tilde{n},\tilde{{\bf
t}},\tilde{{\bf t}}^*,z_{\alpha_i})-\xi(\tilde{n},{\tilde{\bf
t}},{\tilde {\bf t}}^*,z_{-\beta_j})}
\frac{\prod^k_{i<j}(z_{\alpha_i}-z_{\alpha_j})(z_{-\beta_i}-z_{-\beta_j})}
{\prod^k_{i,j}(z_{-\beta_i}-z_{\alpha_j})}\left(\det
a_{\alpha_i\beta_j}|_{i,j=1}^{k}\right) ,
\end{equation}
where the sum ranges all partitions
$(\alpha|\beta)=(\alpha_1,\dots ,\alpha_k|\beta_1,\dots
,\beta_k),\ k=1,2,3,\dots $, and the $\xi(\tilde{n},{\tilde{\bf
t}},{\tilde {\bf t}}^*,z)$ is the function defined by (\ref{xi}).

\quad

Now, compare (\ref{deta''}) and (\ref{KPsoliton'}). We identify
the $\xi$  with the $T$. The factor $\det a_{ij}$ is identified
with the Schur function $s_\lambda({\bf t})$, while the factor
describing the interaction of the solitons is identified with
$s_\lambda({\bf t}^*)$ .  The Lemma 1 yields
\begin{theorem} Let $ \tau(n,{\bf t},T,{\bf t}^*)$ is defined by
(\ref{htf}), and $\tau^{solitons}({\tilde n},\tilde{{\bf
t}},\tilde{{\bf t}}^*)$ is defined by (\ref{inftysol}), where
\begin{equation}
a_{ij}=a_{ij}({\bf t})= s_{(i|j)}({\bf t}) \ ,
\end{equation}
and where, for a given $n$, and for a given set of the numbers
$z_m$, $m=0,\pm 1,\pm 2,\dots$, the variables $\ {\tilde
n},\tilde{{\bf t}},\tilde{{\bf t}}^*$ are related to the variables
$T$ as
\begin{equation}\label{ThTm}
T_{n+m}=\sum_{k=1}^\infty (z_m^k \tilde{t}_k
-z_m^{-k}\tilde{t}^*_k)+{\tilde n}\log z_m+C_{m+n}
\end{equation}
(here the $C_{m+n}$ are constants independent of $\ {\tilde
n},\tilde{{\bf t}},\tilde{{\bf t}}^*$), and
\begin{equation}\label{c+}
c_n =e^{T_{n-1}+\cdots
+T_1+T_0}=e^{\sum_{m=1}^{n}\left(\sum_{k=1}^\infty (z_{-m}^k
\tilde{t}_k -z_{-m}^{-k}\tilde{t}^*_k)+{\tilde n}\log
z_{-m}+C_{n-m}\right) }\ , \quad n
> 0 \ ,
\end{equation}
\begin{equation}\label{c-}
c_n = e^{-T_{n}-\cdots
-T_{-2}-T_{-1}}=e^{-\sum_{m=0}^{-1-n}\left(\sum_{k=1}^\infty
(z_m^k \tilde{t}_k -z_m^{-k}\tilde{t}^*_k)+{\tilde n}\log
z_m+C_{m+n}\right) } \ , \quad n<0 \ .
\end{equation}
and $c_0=1$.
We have\\
(A) If
\begin{equation}
z_m=m+\alpha+1 \ ,\ m\ge 0\ ,\quad z_{-m}=-m+\alpha \ ,\ m>0 \ ,
\end{equation}
\begin{equation}
C_{m+n}=-\log  m! \ ,\ m\ge 0 \ ,\quad C_{m+n}=-\log (-m)!\ ,\ m<0
\ ,
\end{equation}
then
\begin{equation}\label{ThA}
\tau^{solitons}({\tilde n},\tilde{{\bf t}},\tilde{{\bf
t}}^*)e^{-\sum_{k=1}^\infty k{\tilde t}_k{\tilde
t}^*_k}c_n=\tau(n,{\bf t},T,{\bf t}_\infty)
\end{equation}
(B) If
\begin{equation}
z_m=m+\alpha+1 \ , \ m\ge 0 \ ,\quad z_{-m}=-m+\alpha \ ,\ m>0 \ ,
\end{equation}
\begin{equation}
C_{m+n}=\log  \frac{(a)_{m+1}}{m!} \ ,\ m\ge0 \ ,\quad
C_{m+n}=\log
 \frac{(-)^m(a)_{-m}}{(-m)!}\ ,\ m<0 \ ,
\end{equation}
then
\begin{equation}\label{ThB}
\tau^{solitons}({\tilde n},\tilde{{\bf t}},\tilde{{\bf
t}}^*)e^{-\sum_{k=1}^\infty k{\tilde t}_k{\tilde t}^*_k}c_n
=\tau(n,{\bf t},T,{\bf t}(a))
\end{equation}
(C) If
\begin{equation}
z_m=q^{m+1}+b \ , \ m\ge 0 \ ,\quad z_{-m}=q^{-m}+b \ ,\ m>0 \ ,
\end{equation}
\begin{equation}
C_{m+n}= -\log (q;q)_{m}\ ,\ m\ge0 \ ,\quad C_{m+n}=-\log
(q;q)_{-m}\ ,\ m<0 \ .
\end{equation}
then
\begin{equation}\label{ThC}
\tau^{solitons}({\tilde n},\tilde{{\bf t}},\tilde{{\bf
t}}^*)e^{-\sum_{k=1}^\infty k{\tilde t}_k{\tilde t}^*_k}c_n
=\tau(n,{\bf t},T,{\bf t}(\infty,q))
\end{equation}
(D) If
\begin{equation}
z_m=q^{m+1}+b \ , \ m\ge 0 \ ,\quad z_{-m}=q^{-m}+b \ ,\ m>0 \ ,
\end{equation}
\begin{equation}
C_{m+n}=\log \frac {(q^a;q)_{m +1}}{(q;q)_{m}}\ ,\ m\ge 0 \ ,\quad
C_{m+n}=\log \frac {(-)^{m}q^{-(a+1)m }(q^{1-a};q)_{-m}}
{(q;q)_{-m}}\ ,\ m<0 \ ,
\end{equation}
then
\begin{equation}\label{ThD}
\tau^{solitons}({\tilde n},\tilde{{\bf t}},\tilde{{\bf
t}}^*)e^{-\sum_{k=1}^\infty k{\tilde t}_k{\tilde t}^*_k}c_n
=\tau(n,{\bf t},T,{\bf t}(a,q))
\end{equation}

In these relations the $\alpha$ and the $b$ are arbitrary complex
numbers.
\end{theorem}

\begin{remark}
 By the Remark 1, we can restrict sum in (\ref{htf}) over all partitions
 to the subset $S$. In particular, the polynomial tau function of
 type (\ref{htf}) is related to the soliton tau function with a finite
 number of solitons.
\end{remark}
\begin{remark}. We note that the higher times ${\bf t}$ of the KP
hierarchy we started with are integrals of motion for (solitonic)
tau function (\ref{inftysol}) of the second TL hierarchy. And vice
versa: the higher times $ {\tilde n},\tilde{{\bf t}},\tilde{{\bf
t}}^*$ play the role of integrals of motion for the original KP
hierarchy. We therefore call these hierarchies the dual ones.
\end{remark}
\begin{remark}. Let us note that the action of $ SL(2)$ group on the
lattice of the solitonic spectral parameters $\{z_i\ ,\ i=0,\pm
1,\pm 2,\dots \} $:
\begin{equation}\label{sl(2)}
z_i \to \frac{az_i+b}{cz_i+d}
\end{equation}
do not change the factors (\ref{cij}) which describe the
interaction of the solitons. Action (\ref{sl(2)}) affects only
exponents $\xi$ as a certain change of the variables $\tilde{n},
\tilde{\bf t}$ and
 $\tilde{\bf t}^*$.
 (The corresponding infinitesimal action on the tau
function of the dual hierarchy is described in terms of the
Virasoro generators, $L_{-1}=\sum_{m=2}^\infty m{\tilde t}_m
\partial_{\tilde {t}_{m-1}}$ and $L_0,L_{1}$ (see \cite{GO} for details)
and their counterparts ${\bar L}_1,{\bar L}_0,{\bar L}_{-1}$
\cite{T}).
\end{remark}

Let us point out that a different relation between the rational
and the soliton solutions (the duality) was studied in \cite{Mir}. An infinite
number  soliton solutions with the spectral parameters lying on
the lattice appeared in the papers \cite{BL}, \cite{LS} in a
different way and in a different context. In their cases, there is
no the degeneration, described above,  which results in the
appearance of the determinant $\det a_{\alpha_i\beta_j}$ in
(\ref{deta}), which was important to interpret the Schur function
$s_\lambda({\bf t})$.

We now are going to fix the variables ${\bf t}^*$ in a way
generalizing the previous one.

{\bf Multi-component TL}.

A $p$ is a positive integer. We consider the higher times of the
form:
\begin{equation}\label{t^[p]}
{\bf t}^{[p]}=(\underbrace{0,\dots,0,t_1}_{p},
\underbrace{0,\dots,0,t_2}_{p},0,\dots) \ ,\quad {{\bf
t}^*}^{[p]}=(\underbrace{0,\dots,0,t^*_1}_{p},
\underbrace{0,\dots,0,t^*_2}_{p},0,\dots)
\end{equation}

Let us introduce the notations:
\begin{equation}\label{choicetinftyp}
{\bf t}^{[p]}_\infty=(\underbrace{0,\dots,0,1}_{p},0,0,\dots ) \ ,
\end{equation}
\begin{equation}\label{choicet(a)p}
{\bf t}^{[p]}(a)=(\underbrace{0,\dots,0,\frac{a}{1}}_{p},
\underbrace{0,\dots,0,\frac{a}{2}}_{p},0,\dots) \ ,
\end{equation}
\begin{equation}\label{choicetinftyqp}
{\bf t}^{[p]}(\infty,q)=(\underbrace{0,\dots,0,t_1(\infty,q)}_{p},
\underbrace{0,\dots,0,t_2(\infty,q)}_{p},0,\dots) \ ,
\end{equation}
\begin{equation}\label{choicet(a)qp}
{\bf t}^{[p]}(a,q)=(\underbrace{0,\dots,0,t_1(a,q)}_{p},
\underbrace{0,\dots,0,t_2(a,q)}_{p},0,\dots) \ ,
\end{equation}
where $t_m(\infty,q)$ and $t_m(a,q)$ are the same as in
(\ref{choicetinftyq}) and (\ref{choicet(a)q}).

The first output is that by Appendix B we obtain that each of the tau
 functions $\tau_r(n,{\bf
t}^{[p]}$, ${\bf t}^{[p]}_\infty)$, $ \tau_r(n,{\bf t}^{[p]},{\bf
t}^{[p]}(a))$, $\tau_r(n,{\bf t}^{[p]},{\bf t}^{[p]}(\infty,q))$
and  $\tau_r(n,{\bf t}^{[p]},{\bf t}^{[p]}(a,q))$ is just the
product of tau functions we obtained above.

\quad

 Let us consider a set of
partitions $\lambda^{(m)}=(\alpha_1^{(m)},\alpha_2^{(m)},
\dots,\alpha_{k^{(m)}}^{(m)} |\beta_1^{(m)},\beta_2^{(m)},\dots
,\beta_{k^{(m)}}^{(m)})$, where $ m=1,\dots,p$ and $k^{(1)}+\cdots
+k^{(p)} =k$, where $p$ and $k$ are two positive integers.
Consider the set of numbers $\{\alpha_{k_i}^{(m)}p+m-1
|i=1,\dots,k^{(m)}, m=1,\dots,p\}$ and put them in the decreasing
order: $\alpha_1>\alpha_2>\cdots
>\alpha_k>0$ (by the construction of
numbers $\alpha_k$ the decreasing order is strict ). Similarly
consider the set  $\{\beta_{k_i}^{(m)}p+m-p |i=1,\dots,k^{(m)},
m=1,\dots,p\}$ and put them in the decreasing order:
$\beta_1>\beta_2>\cdots
>\beta_k>0$. Let us
consider the partition $\lambda=(\alpha_1,\dots,\alpha_k |\beta_1,
\dots,\beta_k)$. We can prove the following lemma.

\begin{Lemma}. For the partitions
$(\alpha^{(m)}|\beta^{(m)})=$ $(\alpha_1^{(m)},\dots
,\alpha_k^{(m)} |\beta_1^{(m)},\dots ,\beta_k^{(m)})$, where $
m=1,\dots,p$ and the partition $(\alpha|\beta)=(\alpha_1,\dots
,\alpha_k|\beta_1,\dots ,\beta_k)$  described above, we have
\begin{equation}\label{stinfty[p]}
s_{(\alpha|\beta)}({\bf
t}_\infty^{[p]})=(-)^{\sigma(\alpha,\beta)}\prod^p_{m=1}
\frac{\prod^{k^{(m)}}_{i<j}(\alpha_i^{(m)}-\alpha_j^{(m)})
(\beta_i^{(m)}-\beta_j^{(m)})}
{\prod_{i,j=1}^k(\alpha_i^{(m)}+\beta_j^{(m)}+1)}\frac
{1}{\prod_{i=1}^{k^{(m)}} \alpha_i^{(m)}! \beta_i^{(m)}!} \ ,
\end{equation}
\begin{equation}\label{st(a)[p]}
s_{(\alpha|\beta)}({\bf
t}^{[p]}(a))=(-)^{\sigma(\alpha,\beta)}
\prod^p_{m=1}\frac{\prod^{k^{(m)}}_{i<j}
(\alpha_i^{(m)}-\alpha_j^{(m)})(\beta_i^{(m)}-\beta_j^{(m)})}
{\prod_{i,j=1}^k(\alpha_i^{(m)}+\beta_j^{(m)}+1)}
\prod_{i=1}^{k^{(m)}}\frac {(a)_{\alpha_i^{(m)}+1}}{\alpha_i^{(m)}!}
\frac{(-1)^{\beta_i^{(m)}}(-a)_{\beta_i^{(m)}}}{
\beta_i^{(m)}!} \ ,
\end{equation}
\begin{equation}\label{stinftyq[p]}
s_{(\alpha|\beta)}({\bf
t}^{[p]}(\infty,q))=
\end{equation}
\begin{equation}\label{stinftyq'[p]}
(-)^{\sigma(\alpha,\beta)}\prod^p_{m=1}
\frac{\prod^{k^{(m)}}_{i<j}(q^{\alpha_i^{(m)}+1}-q^{\alpha_j^{(m)}+1})
(q^{-\beta_j^{(m)}}-q^{-\beta_i^{(m)}})}
{\prod_{i,j=1}^k(q^{-\beta_i^{(m)}}-q^{\alpha_j^{(m)}+1})}\frac
{1}{\prod_{i=1}^{k^{(m)}}(q;q)_{\alpha_i^{(m)}}
(q;q)_{\beta_i^{(m)}}} \ ,
\end{equation}
\begin{equation}\label{st(a,q)[p]}
s_{(\alpha|\beta)}({\bf
t}^{[p]}(a,q))=(-)^{\sigma(\alpha,\beta)}
\prod^p_{m=1}\frac{\prod^{k^{(m)}}_{i<j}
(q^{\alpha_i^{(m)}+1}-q^{\alpha_j^{(m)}+1})
(q^{-\beta_j^{(m)}}-q^{-\beta_i^{(m)}})}
{\prod_{i,j=1}^{k^{(m)}}(q^{-\beta_i^{(m)}}-q^{\alpha_j^{(m)}+1})}
\end{equation}
\begin{equation}\label{st(a,q)[p]'}
\cdot \prod_{i=1}^{k^{(m)}}\frac {(q^a;q)_{\alpha_i^{(m)}
+1}}{(q;q)_{\alpha_i^{(m)}}}\frac
{(-1)^{\beta_i^{(m)}}q^{(a-1)\beta_i^{(m)}
}(q^{1-a};q)_{\beta_i^{(m)}}} {(q;q)_{\beta_i^{(m)}}}
\end{equation}
\end{Lemma}

{\bf Soliton solutions of multi-component TL ($p$-TL)}. The
following expression is an example of tau function of the
multi-component TL:
\begin{equation}\label{mTLtau'}
\tau^{solitons}(\tilde{n}^{(1)},\dots,\tilde{n}^{(p)};{\bf \tilde
{ t}}^{(1)},\dots,{\bf \tilde{t}}^{(p)};{\bf
{\tilde{t}}}^{*(1)},\dots,{\bf \tilde{t}}^{*(p)})=
\end{equation}
\begin{equation}\label{mTLtau}
\l {\tilde n}|e^{\sum_{m=1}^p\sum_{k=1}^\infty H^{(m)}_k{\tilde
t}_k^{(m)}} e^{\sum_{m,n=1}^p\sum_{i\ge0,j>0}^\infty
a_{ij}^{mn}{\psi}^{(m)}(z_{i,m}^{(m)}){\psi^*}^{(n)}(z_{-j,n})}
e^{\sum_{m=1}^p\sum_{k=1}^\infty H^{(m)}_{-k}{{{\tilde
t}_{-k}}}^{(m)} }|{\tilde n}\r
\end{equation}
where the sets $({\tilde t}_1^{(m)}, {\tilde t}_2^{(m)},\dots)\ ,\
({{{\tilde t}_{-1}}}^{(m)},{{{\tilde
t}_{-2}}}^{(m)},\dots),m=1,\dots,p$ are higher times of
multi-component TL hierarchy and
\begin{equation}\label{tilden}
{\tilde n} =({\tilde n}^{(1)}, \dots, {\tilde n}^{(p)})
\end{equation}
 are $p$-TL discrete variables such that ${\tilde
n}^{(1)}+ \cdots +{\tilde n}^{(p)}=0$, and where
\begin{equation}\label{H^mk}
H^{(m)}_k=\sum_{n=-\infty}^{+\infty}\psi_{n}^{(m)} {\psi^*_{n+k}}^{(m)} \ ,
k \neq 0 \ ,
\end{equation}
\begin{equation}\label{psim}
\psi_n^{(m)}:=\psi_{np+m-1} \ ,\quad
{\psi^*_n}^{(m)}:=\psi^*_{np+m-1} \ ,
\end{equation}
\begin{equation}\label{mfields}
{\psi}^{(m)}(z):=\sum_{k=-\infty}^\infty z^{k}\psi_k^{(m)} \
,\quad {\psi^*}^{(m)}(z)dz:= \sum_{k=-\infty}^\infty
z^{-k-1}{\psi^*_k}^{(m)}
\end{equation}
This soliton tau function is defined by numbers $a_{ij}^{mn}\
(i\ge 0,j>0, n,m=1,\dots,p)$ and the set of $z_{j,m}$ where
$m=1,\dots,p$ and $j$ runs over all integers.

We have the generalization of the Theorem 1.

\begin{theorem} Let
$ \tau^{solitons}(\tilde{n}^{(1)},\dots,\tilde{n}^{(p)};{\bf
\tilde { t}}^{(1)},\dots,{\bf \tilde{t}}^{(p)};{\bf
{\tilde{t}}}^{*(1)},\dots,{\bf \tilde{t}}^{*(p)})$ is defined by
(\ref{mTLtau}) and
\begin{equation}
a_{ij}^{mn}=a_{ij}^{mn}({\bf t})= s_{(ip+m-1|jp+n-1)}({\bf t}) \ ,
\end{equation}
and
\begin{equation}\label{ThTmp}
T_k^{(m)}=T_{pk+m-1} \ ,\quad m=1,\dots,p \ ,
\end{equation}
\begin{equation}\label{ThTmp'}
T_k^{(m)}=\sum_{n\neq 0}^\infty (z_{k,m})^n \tilde{t}_n^{(m)}
-{\tilde n}^{(m)}\log z_{k,m}+C_{k}^{(m)}\ ,\ k\in Z \ ,
\end{equation}
where $C_{k}^{(m)}$ are some constants, specified below

Then\\
(A)
\begin{equation}\label{ThAp}
\tau^{solitons}(\tilde{n}^{(1)},\dots,\tilde{n}^{(p)};{\bf \tilde
{ t}}^{(1)},\dots,{\bf \tilde{t}}^{(p)};{\bf
{\tilde{t}}}^{*(1)},\dots,{\bf
\tilde{t}}^{*(p)})e^{-\sum_{m=1}^p\sum_{n=1}^\infty n{\tilde
t}_n^{(m)}{\tilde t}^{*(m)}_n}=\tau\left(n,{\bf t},T,{\bf
t}^{[p]}_\infty \right)
\end{equation}
where the l.h.s. describe the $\infty$-soliton solution, with
 the parameters:
\begin{equation}
z_{k,m}=k+\alpha^{(m)}+1 \ ,\ k\ge 0\ ,\quad
z_{-k,m}=-k+\alpha^{(m)} \ ,\ k>0 \ , \quad m=1,\dots,p \ ,
\end{equation}
\begin{equation}
C_{k}^{(m)}=-\log  k! \ ,\ k\ge 0 \ ,\quad C_{k}^{(m)}=-\log
(-k)!\ ,\ k<0 \ .
\end{equation}
(B)
\begin{equation}\label{ThBp}
 \tau^{solitons}(\tilde{n}^{(1)},\dots,\tilde{n}^{(p)};{\bf
\tilde { t}}^{(1)},\dots,{\bf \tilde{t}}^{(p)};{\bf
{\tilde{t}}}^{*(1)},\dots,{\bf
\tilde{t}}^{*(p)})e^{-\sum_{m=1}^p\sum_{n=1}^\infty n{\tilde
t}_n^{(m)}{\tilde t}^{*(m)}_n}= \tau(n,{\bf t},T,{\bf t}^{[p]}(a))
\end{equation}
where the l.h.s. describe the $\infty$-soliton solution with the
parameters
\begin{equation}
z_{k,m}=k+\alpha^{(m)}+1 \ , \ k\ge 0 \ ,\quad
z_{-k,m}=-k+\alpha^{(m)} \ ,\ k>0 \ ,
\end{equation}
\begin{equation}
C_{k}^{(m)}=\log  \frac{(a)_{k+1}}{k!} \ ,\ k\ge0 \ ,\quad
C_{k}^{(m)}=\log
 \frac{(-)^k(a)_{-k}}{(-k)!}\ ,\ k<0 \ .
\end{equation}
(C)
\begin{equation}\label{ThCp}
 \tau^{solitons}(\tilde{n}^{(1)},\dots,\tilde{n}^{(p)};{\bf
\tilde { t}}^{(1)},\dots,{\bf \tilde{t}}^{(p)};{\bf
{\tilde{t}}}^{*(1)},\dots,{\bf
\tilde{t}}^{*(p)})e^{-\sum_{m=1}^p\sum_{n=1}^\infty n{\tilde
t}_n^{(m)}{\tilde t}^{*(m)}_n}=\tau(n,{\bf t},T,{\bf
t}^{[p]}(\infty,q))
\end{equation}
where the l.h.s. describe the $\infty$-soliton solution with the
parameters
\begin{equation}
z_{k,m}=q^{k+1}+b^{(m)} \ , \ k\ge 0 \ ,\quad
z_{-k,m}=q^{-k}+b^{(m)} \ ,\ k>0 \ ,
\end{equation}
\begin{equation}
C_{k}^{(m)}= -\log (q;q)_{k}\ ,\ k\ge0 \ ,\quad C_{k}^{(m)}=-\log
(q;q)_{-k}\ ,\ k<0 \ .
\end{equation}
(D)
\begin{equation}\label{ThDp}
 \tau^{solitons}(\tilde{n}^{(1)},\dots,\tilde{n}^{(p)};{\bf
\tilde { t}}^{(1)},\dots,{\bf \tilde{t}}^{(p)};{\bf
{\tilde{t}}}^{*(1)},\dots,{\bf \tilde{t}}^{*(p)})
e^{-\sum_{m=1}^p\sum_{n=1}^\infty n{\tilde t}_n^{(m)}{\tilde
t}^{*(m)}_n} =\tau(n,{\bf t},T,{\bf t}^{[p]}(a,q))
\end{equation}
where the l.h.s. describe the  $\infty$-soliton solution with the
parameters
\begin{equation}
z_{k,m}=q^{k+1}+b^{(m)} \ , \ k\ge 0 \ ,\quad
z_{-k,m}=q^{-k}+b^{(m)} \ ,\ k>0 \ , \quad
\end{equation}
\begin{equation}
T_{k,0}^{(m)}=\log \frac {(q^a;q)_{m +1}}{(q;q)_{m}}\ ,\ m\ge 0 \
,\quad T_{k,0}^{(m)}=\log \frac {(-)^{k}q^{-(a+1)k
}(q^{1-a};q)_{-k}} {(q;q)_{-k}}\ ,\ k<0
\end{equation}
In the relations above $\alpha^{(m)}$ and $b^{(m)}$ are arbitrary
complex numbers, and ${\bf t}^{[p]}_\infty,{\bf t}^{[p]}(a),{\bf
t}^{[p]}(\infty,q)$ and ${\bf t}^{[p]}(a,q)$ are specified
respectively by
(\ref{choicetinftyp}),(\ref{choicet(a)p}),(\ref{choicetinftyqp})
and (\ref{choicet(a)qp}).
\end{theorem}

\section{ Matrix models as soliton solutions}

Let us describe without details the relation of asymptotic series
for certain matrix models to the $\infty$-soliton solutions which
are the subject of the Theorem 1. More details about asymtotic
series of matrix models one can find in \cite{Cadiz},\cite{Or}.

\subsection{Normal matrix model}

A good example of tau function of hypergeometric type (\ref{htf})
is the asymptotic perturbation series of the model of normal
matrices with an axial symmetric interaction term (this term
defines the set of variables $T$ in (\ref{htf})). The model of
normal matrix has a set of interesting applications, mentioned
below.

A matrix $M$ is called {\em normal} if it commutes with its
hermitian conjugated $M^+$:
\begin{equation}
[M,M^+]=0
\end{equation}

  One can bring the matrix $M$ to
its diagonal form via an unitary matrix $U$: $M=UZU^+$, where
$Z=diag(z_1,\dots,z_n)$, the $z_i$ are eigenvalues of the $M$. It
is clear that $M^+=U{\bar Z}U^+$, where the diagonal matrix ${\bar
Z}$ is the complex conjugated of $Z$.

 The model of normal matrices is defined by its partition function as
 follows
\begin{equation}\label{normint}
Z(n,{\bf t},{\bf t^*};V)=\int d\Omega(M)
e^{\textrm{Tr}V_1(M)+\textrm{Tr}V_2(M^+)+\textrm{Tr}V(M,M^+)}
\end{equation}
The integration measure is defined as
\begin{equation}\label{normmeasure}
d\Omega(M)=d_*U|\Delta(z)|^2\prod_{i=1}^n d^2z_i
\end{equation}
where $d_*U$ is the Haar measure over the unitary group
$\textrm{U}(n)$ and $\Delta(z)$ is the notation for the Vandermont
determinant:
\begin{equation}\label{van}
\Delta(z)=\det z_i^{n-k}|_{i,k=1,\dots,n}=\prod_{i<k}^n(z_i-z_k)
\end{equation}
For the case $n=1$ one puts $\Delta(z)=1$.

 The potentials
$V_1(M),V_2(M^+)$ and $V(M,M^+)$ are defined by their Taylor
expansion as follows
\begin{equation}\label{UV}
V_1(M)=\sum_{m=1}^\infty t_m  M^m,\quad V_2(M)=\sum_{m=1}^\infty
t_m^* \left(M^+\right)^m \ ,\quad V(M,M^+)=\sum_{k,m=1}^\infty
v_{km} M^k(M^+)^m \ ,
\end{equation}
where the $t_m ,t_m^*,v_{km}  $ are complex numbers.

After the change of variables $M\to (U,Z)$ and the integration
over $\textrm{U}(n)$  one obtains
\begin{equation}\label{normint-}
Z(n,{\bf t},{\bf t^*};V)=C\int_{\texttt c} \cdots \int_{\texttt c}
|\Delta(z)|^2 \prod_{i=1}^n e^{V_1(z_i)+V_2({\bar
z}_i)+V(z_i,{\bar z}_i)}d^2z_i \ ,
\end{equation}
where  the integration is going over complex planes of eigenvalues
$z_i$. $C$ is a number related to the volume of the unitary group.

The model of normal matrices has applications in the description
of quantum Hall droplets. It was shown in \cite{MWZ}, the
particular case of the model when
$v_{km}=\delta_{k,1}\delta_{m,1}$, in its $1/n\to \infty$ limit is
related to the interface dynamics of a water spot inside an oil
film, and that it is also related to certain old posed problems of
complex analysis.

Now, we derive the perturbative expansion of partition function
(\ref{normint-}) in the coupling constants ${\bf t}$ and ${\bf
t}^*$. We set
\begin{equation}\label{eUneq}
e^{T_0-T_m}=A \int_0^{+\infty} z^me^{\sum_{k=1}^\infty v_kz^k} dz
\end{equation}
Then, according to \cite{Or}, we have the following asymptotic
series:
\begin{equation}\label{KPsoliton'NM}
I^{NM}(n,{\bf t},{\bf t^*};V)=\tau_r(n,{\bf t},{\bf
t}^*)=I^{NM}(n,0,0;V)\sum_{(\alpha|\beta)}s_{(\alpha|\beta)}({\bf
t})s_{(\alpha|\beta)}({\bf t}^*) e^{\sum_{i=1}^{k(\alpha|\beta)}
(T_{n-\beta_i-1}-T_{n+\alpha_i})} \ ,
\end{equation}
which, for the choice of the times ${\bf t}^*$, according to
(\ref{choicetinftyp}),(\ref{choicet(a)p}),(\ref{choicetinftyqp}),
(\ref{choicet(a)qp}),  can be interpreted as the infinite-soliton
solution of the Theorems 1,2.

For instance, on the level of perturbation series the matrix
integral
\begin{equation}\label{normintp}
I^{NM}(n,{\bf t}^{(p)}(a),{\bf t^*}^{(p)}(a);V)=\int d\Omega(M)
e^{\textrm{Tr}U(MM^+)}\det(1-M^p)^{-a}\det(1-(M^+)^p)^{-a}
\end{equation}
factorizes into the product of the $p$ tau functions of the
$\infty$-soliton type.

\subsection{Two-matrix model}

Let us evaluate the following integral over $n$ by $n$  matrices
$M_1$ and $M_2$, where $M_1$ is a Hermitian matrix and $M_2$ is an
anti-Hermitian one
\begin{equation}\label{HTMMmeasure}
I^{2MM}(n,{\bf t},{\bf t^*})=\int
e^{\textrm{Tr}V_1(M_1)+\textrm{Tr}V_2(M_2)}e^{-\textrm{Tr}
M_1M_2}dM_1dM_2 \ ,
\end{equation}
where the integration measure is
\begin{equation}\label{measure2MM}
dM_1dM_2 = \prod_{i=1}^n d(M_1)_{ii}d(M_2)_{ii}\prod_{i<j}^n
d\Re(M_1)_{ij} d\Re(M_2)_{ij}d\Im(M_1)_{ij} d\Im(M_2)_{ij}
\end{equation}
and
\begin{equation}\label{HTMMV}
V_1(M_1)=\sum_{m=1}^\infty t_m M_1 ^m \ ,\quad
V_2(M_2)=\sum_{m=1}^\infty t_m^* M_2^m
\end{equation}
It is well-known \cite{IZ} that this integral reduces to the
integral over eigenvalues $x_i$ and $y_i$ of matrices $M_1$ and
$M_2$ respectively.

We have \cite{Cadiz}
\begin{equation}\label{2MM=scalprod}
I^{2MM}(n,{\bf t},{\bf
t^*})=\frac{C_n}{(2\pi)^n}\sum_{\lambda}(n)_\lambda s_\lambda({\bf
t})s_\lambda({\bf t^*}) \ ,
\end{equation}
where we remember that according to (\ref{rlambda})
\begin{equation}\label{(n)lambda}
(n)_\lambda=\prod_{i,j\in
\lambda}(n+j-i)=\frac{\Gamma(n+1+\lambda_1)\Gamma(n+\lambda_2)\cdots
\Gamma(\lambda_n)}{\Gamma(n+1)\Gamma(n)\cdots \Gamma(1)} \ ,
\end{equation}
and $C_n$ incorporates the volume of unitary group.

Thus we have the following perturbation series \cite{Cadiz}
\begin{equation}\label{2MMss}
\frac{I^{2MM}(n,{\bf t},{\bf t^*})}{I^{2MM}(n,0,0)}
=\sum_{\lambda} (n)_{\lambda}s_{\lambda}({\bf t})s_{\lambda}({\bf
t}^*)
\end{equation}
It means that for the choice of times, ${\bf t}^*$, according to
(\ref{choicetinftyp}),(\ref{choicet(a)p}),(\ref{choicetinftyqp}),
or (\ref{choicet(a)qp}), partition function (\ref{HTMMmeasure})
can be interpreted as the infinite-soliton solution of the Theorem
1.

For instance, on the level of perturbation series, the matrix
integral
\begin{equation}\label{HTMMmeasurep}
I^{2MM}(n,{\bf t}^{(p)}_\infty,{\bf t}^{(p)}_\infty)=\int
e^{\textrm{Tr}M_1^p+\textrm{Tr}M_2^p}e^{-\textrm{Tr}
M_1M_2}dM_1dM_2
\end{equation}
factorizes into the product of $p$ tau functions of the
$\infty$-soliton type.

\subsection{Hermitian one matrix model}

The partition function for the one matrix model with an even
potential is
\begin{equation}\label{1HOMM}
Z(N,g,g_4)=\int dM e^{-N\frac g2\textrm{Tr}
M^2-N\sum_{m>2}\frac{g_{2m}}{2m}\textrm{Tr} M^{2m}} \ ,
\end{equation}
where $M$ is $N$ by $N$ Hermitian matrix, and $dM = \prod_{i=1}^n
d M_{ii}\prod_{i<j}^n  d\Re M_{ij}d\Im M_{ij}$.

 It follows from \cite{Cadiz} and from Appendix B, that the
asymptotic series of the one matrix model with the even potential
enjoys the factorization property:
\begin{equation}\label{difexpression}
Z(2N,g,g_4)=\tau_{r^{(1)}}(N,{\bf t},{\bf
t}^*)\tau_{r^{(2)}}(N,{\bf t},{\bf t}^*) \ ,
\end{equation}
\begin{equation}\label{difexpression'}
Z(2N+1,g,g_4)=\tau_{r^{(1)}}(N+1,{\bf t},{\bf
t}^*)\tau_{r^{(2)}}(N,{\bf t},{\bf t}^*) \ ,
\end{equation}
where
\begin{equation}\label{tt*1MMdiff}
{\bf t}=\left(0,-Ng_4/4,-Ng_6/6,-Ng_8/8,\dots \right),\quad {\bf
t}^*=\left(1/(2Ng),0,0,0,\dots \right)=1/(2Ng){\bf t}_\infty
\end{equation}
Therefore, each factor in (\ref{difexpression}) is the subject of
(\ref{ThA}) and has the form of an $\infty$-soliton solution.

In (\ref{difexpression}),(\ref{difexpression'})
\begin{equation}\label{rdiff}
r^{(1)}(k)=2k(2k-1) ,\quad r^{(2)}(k)=(2k+1)(2k) \ ,
\end{equation}
which defines $T_m$ in (\ref{KPsoliton'}) for the tau functions
$\tau_{r^{(1)}}(N+1,{\bf t},{\bf t}^*)$ and $\tau_{r^{(2)}}(N,{\bf
t},{\bf t}^*)$ via (\ref{rT}).

 Remark. In case $t^*_k=\delta_{k,4}$, we note that  the series for
the $\tau_{r^{(1,2)}}(N,{\bf t},{\bf t}^*)$ can be obtained as the
perturbation series for the respective partition functions
\begin{equation}\label{1MM=ZZ}
\tau_{r^{(1,2)}}(N,{\bf t},{\bf t}^*)=C_N\int
e^{-\frac{\textrm{Tr}{M^2}}{2Ng_4}}\det
\left(1-\frac{\sqrt{-2}M}{gN}\right)^{-N\pm\frac 12}dM
\end{equation}
Here, the integration ranges Hermitian $N$ by $N$ matrices $M$,
and $C_N$ is the normalization constant. Thus, the asymptotic
series for integral (\ref{1MM=ZZ}) has the form of an
$\infty$-soliton tau function.

\section{Appendix A. Toda lattice tau functions of
hypergeometric type \cite{nl64},\cite{pd22}}

We shall use the fermionic representation of the paper \cite{DJKM}
and \cite{JM} (see also \cite{UT}).

 The KP tau function of the
hypergeometric type has the fermionic representation (this result
is indicated in \cite{nl64}; the details of the proof can be found
in \cite{Cadiz}),
\begin{equation}\label{tauhyp1}
\tau_r(n,{\bf t},{\bf t^*} )\ =\langle n|e^{H({\bf t})} e^{-A(
{\bf t^* })} |n\rangle \ ,
\end{equation}
which a special case of the fermionic representation of the KP tau
function \cite{JM}, where
\begin{equation}
H({\bf t})=\sum_{m=1}^\infty H_mt_m \ , \quad A( {\bf t^* })=
\sum_{m=1}^\infty A_mt_m^* \ ,
\end{equation}
\begin{equation}\label{H_m}
H_m= \sum_{k=-\infty}^\infty \psi_k\psi^*_{k+m} \
 , \quad m\neq 0 \ ,
\end{equation}
\begin{equation}\label{A_m}
A_m= -\sum_{n=-\infty}^\infty r(n)\cdots
r(n-m+1)\psi_n\psi^*_{n-m} \
 , \quad m=1,2,\dots
\end{equation}
Here, $\{\psi_m\}, \{\psi_m^*\}$ (${m \in {\bf Z}}$) are the
fermionic operators, which satisfy the usual anti-commutation
relations
\begin{equation}\label{fermionscomrel}
[\psi_j, \psi_k^*]_+ = \delta_{jk}, \quad [\psi_j, \psi_k]_+=0,
\quad [\psi_j^*, \psi_k^*]_+ = 0, \quad  j,k\in {\bf Z} \ ,
\end{equation}
The vacuum  state of a charge $n$ in (\ref{tauhyp1}) is defined as
\begin{equation}\label{nvac}
|n\r := \psi_{n-1} \cdots \psi_1 \psi_0 | 0\r \ ,\ n>0\ ,\qquad
|n\r := \psi^*_{n} \cdots \psi^*_{-1} | 0\r \ ,\ n<0\ ,
\end{equation}
where the vacuum state $|0\r$  is characterized by
\begin{equation}\label{vac0}
\psi_m|0\r =0, \quad m<0; \qquad \psi^*_m |0\r =0,\quad m \ge 0
\end{equation}

Due to (\ref{H_m}) and (\ref{A_m}), we have two different abelian
group actions, respectively generated  by the set
$\{H_n,n=1,2,\dots\}$ and by the set $\{A_n,n=1,2,\dots\}$.
However, operators from different sets do not commute with each
other. We also define
\begin{equation}\label{H*}
H^*({\bf t}^*)=\sum_{m=1}^{\infty}H_{-m}t_m^* \ ,
\end{equation}
where $H_m$ are given in (\ref{H_m}). Then, we have  $A({\bf
t}^*)=H^*({\bf t}^*)$ for $r\equiv 1$.

We also recall  the so-called {\bf Wick rule} , which is useful in
calculations
\begin{equation}\label{Wick}
\l n|w_1 \cdots w_{2m+1}|n \r =0 \ ,\quad \l n|w_1 \cdots w_{2m}
|n\r =\sum_\sigma sgn\sigma \l n|w_{\sigma(1)}w_{\sigma(2)}|n\r
\cdots \l n| w_{\sigma(2m-1)}w_{\sigma(2m)} |n\r \ ,
\end{equation}
where $w_k,k=1,2,\dots $ is any liner combination of the fermionic
operators $\psi_i$ and $\psi_i^*$ ($i=0,\pm 1,\pm 2,\dots$), and
where $\sigma$ ranges all permutations satisfying
$\sigma(1)<\sigma(2),\dots , \sigma(2m-1)<\sigma(2m)$ and
$\sigma(1)<\sigma(3)<\cdots <\sigma(2m-1)$ (here, $\sigma(k)$ is
the result of the action of the element $\sigma$ of the
permutation group on the number $k$)

These linear combinations
\begin{equation}\label{fields}
\psi(z)=\sum_{m=-\infty}^{+\infty} \psi_m z^m \ ,\quad
\psi(z)^*=\sum_{m=-\infty}^{+\infty} \psi_m^* z^{-m-1}
\end{equation}
are of most use. By (\ref{Wick}) we have
\begin{equation}\label{fervan}
\l n|\psi(z_1)\cdots\psi(z_k)
\psi^*(z_{-1})\cdots\psi^*(z_{-k})|n\r=\left( \frac{z_1}{z_{-1}}
\right)^n\cdots \left( \frac{z_k}{z_{-k}}
\right)^n\frac{\prod_{i<j}^k(z_i-z_j)(z_{-i}-z_{-j})}
{\prod_{i,j=1}^k(z_{i}-z_{-j})}
\end{equation}

We have the simple relations \be\label{xit}
 e^{H({\bf t})}\psi(z) e^{-H({\bf t})}= \psi(z) e^{\xi({\bf t},z)},
\quad e^{H({\bf t})}\psi^*(z) e^{-H({\bf t})} = \psi^*(z)
e^{-\xi({\bf t},z)} , \ee \be\label{xit*}
 e^{-H^*({\bf t}^*)}\psi(z) e^{H^*({\bf t}^*)}=
\psi(z) e^{-\xi({\bf t}^*,z^{-1})}, \quad e^{-H^*({\bf
t}^*)}\psi^*(z) e^{H^*({\bf t}^*)}= \psi^*(z) e^{\xi({\bf
t}^*,z^{-1})}  \ee where
\begin{equation}\label{xiHH*}
\xi({\bf t},z)=\sum_{m=1}^\infty z^m t_m
\end{equation}

We now consider {\em Toda lattice} (TL) \cite{DJKM},\cite{UT}. Our
notations $n,{\bf t},{\bf t}^*$ correspond to the respective
notations $s,x,-y$ in \cite{UT}.

We consider TL tau-function (\ref{htf}), which depends on the
three sets of variables ${\bf t},{T},{\bf t}^*$ and on $n \in Z$:
\begin{equation}\label{tau3}
\tau(n,{\bf t},{T},{\bf t}^*)= \langle n|e^{H({\bf t})}
\exp\left(\sum_{-\infty}^{\infty}T_m:\psi_m^*\psi_m:\right)
e^{H^*({\bf t}^*)}|n\rangle \ ,
\end{equation}
where $:\psi_m^*\psi_m:=\psi^*_m\psi_m -\l 0| \psi_m^*\psi_m |0\r
$.  With respect to the KP and the TL dynamics, the variables
$T_n$ have a meaning of integrals of motion.

Hypergeometric functions
(\ref{ordinary}),(\ref{qordinary}),(\ref{vh}),(\ref{hZ1}) listed
in the Appendix are ratios of tau-functions (\ref{tau3}) evaluated
at special values of times $n,{\bf t},{T},{\bf t^*}$
\cite{nl64},\cite{pd22}.

Let us mark that tau functions (\ref{tau3}) with special choice of
$T$ variables were considered in \cite{T},\cite{NTT} in the
context of $c=1$ string theory.

Tau-function (\ref{tau3}) is linear in each $e^{T_n}$.
 Let $r(m)=e^{T_{m-1}-T_m}$, $m\in Z$. Then
\begin{equation}\label{tauhyp'}
\tau_r(n,{\bf t},{\bf t}^*)=c_n(T)^{-1}\tau (n,{\bf t},{T},{\bf
t}^* )=\frac{\tau (n,{\bf t},{T},{\bf t}^* )} {\tau (n,{\bf
0},{T},{\bf 0} )}=
\end{equation}
\begin{equation}\label{tauhyp''}
 1+ \sum_{{\lambda}\neq {\bf 0}}
e^{(T_{n-1}-T_{\lambda_1+n-1})+(T_{n-2}-T_{\lambda_2+n-2})+\cdots
+ (T_{n-l}-T_{\lambda_l+n-l}) }s_{\lambda }({\bf t}) s_{\lambda
}({\bf t }^*)
\end{equation}
The sum ranges all different partitions
excluding the partition ${\bf 0}$.\\

 We have
\begin{equation}\label{H0+}
c_n^{-1}=\tau (n,{\bf 0},{T},{\bf 0} ) =e^{-T_{n-1}-\cdots
-T_1-T_0}=e^{-nT_0}\prod_{k=0}^{n-1}\left(r(k)\right)^{n-k}\ ,
\quad n
> 0 \ ,
\end{equation}
\begin{equation}\label{H00}
c_n^{-1}=\tau (0,{\bf 0},{T},{\bf 0} )=1 \ , \quad n=0 \ ,
\end{equation}
\begin{equation}\label{H0-}
c_n^{-1}=\tau (n,{\bf 0},{T},{\bf 0} ) = e^{T_{n}+\cdots
+T_{-2}+T_{-1}}=e^{-nT_0}\prod_{k=0}^{-n-1}\left(r(k)\right)^{-k-n}
\  , \quad n<0 \ .
\end{equation}

The KP and TL tau-function $\tau(n)=\tau_r(n,{\bf t},{\bf t}^*)$
obeys the following Hirota equation:
\begin{equation}\label{hirota}
\tau(n)\partial_{t^*_1}
\partial_{t_1}\tau(n)-
\partial_{t_1}\tau(n)
\partial_{t^*_1}\tau(n)=r(n)\tau(n-1)\tau(n+1)
\end{equation}

The equation
\begin{equation}\label{rToda}
\partial_{t_1}\partial_{t^*_1}\phi_n=
r(n)e^{\phi_{n-1}-\phi_{n}}- r(n+1)e^{\phi_{n}-\phi_{n+1}} \ ,
\end{equation}
which is similar to the Toda lattice equation holds for
\begin{equation}\label{phi_n}
\phi_n({\bf t},{\bf t}^*)=-\log \frac{\tau_r(n+1,{\bf t},{\bf
t}^*)}{\tau_r(n,{\bf t},{\bf t}^*)}
\end{equation}

Equations (\ref{rToda}) and (\ref{hirota}) are still true in case
the function $r$ has zeroes.\\
If the function $r$ has no integer zeroes, using the change of
variables
\begin{equation}\label{phivarphi}
 \varphi_n=-\phi_n - T_n \ ,
\end{equation}
we obtain Toda lattice equation in the standard form \cite{UT}:
\begin{equation}\label{Toda}
\partial_{t_1}\partial_{t^*_1}\varphi_n=e^{\varphi_{n+1}-\varphi_{n}}-
e^{\varphi_{n}-\varphi_{n-1}}
\end{equation}
As we see from (\ref{phivarphi}), the variables $T_n$ have the
meaning of asymptotic values of the fields $\phi_n$ for the class
of tau-functions (\ref{tau3}) which is characterized by the
property $\varphi_n \to 0$ as $t_1 \to 0$.

\section{Appendix B. Multi-component case}

\bprop Let $p$ be a positive integer.  Given $i=1,2,\dots ,p$, we
have
\begin{equation}\label{tauhyp1tN}
\tau_r(np+i,{\bf t}^{[p]},{\bf t^*}^{[p]} )
=\tau_{r^{(1)}}(n+1,{\bf {t}},{\bf {t}^*} )\cdots
\tau_{r^{(i)}}(n+1,{\bf {t}},{\bf {t}^*} )\tau_{r^{(i+1)}}(n,{\bf
{t}},{\bf {t}^*} )\cdots \tau_{r^{(p)}}(n,{\bf {t}},{\bf {t}^*} )\
,
\end{equation}
where $i=1,\dots , p$, and each $\tau_{r^{(m)}}(j,{\bf {t}},{\bf
{t}^*} )$ (with $j=n,n+1$) is given by (\ref{tauhyp1}) with the
function $r$, defined by
\begin{equation}\label{tNr_i}
r^{(m)}(k)={R}(kp+m-1) \ ,\quad m=1,\dots , p,\quad k \in Z \ ,
\end{equation}
where
\begin{equation}\label{Rr}
R(k)=r(k)r(k-1)\cdots r(k-p+1) \ ,
\end{equation}
and where the times ${\bf t}^{[p]},{\bf t^*}^{[p]} $ in the l.h.s.
 are related to the times
\begin{equation}\label{tildett*}
{\bf {t}}=(t_1,t_{2},t_{3},\dots) \ ,\quad {\bf
{t}^*}=(t_1^*,t_{2}^*,t_{3}^*,\dots)
\end{equation}
in the r.h.s. by (\ref{t^[p]}). \eprop
 Proof. Following \cite{JM}, we introduce the
 {\em colored fermions} according
to $\psi^{(m)}_k:=\psi_{kp+m-1},{\psi^*}^{(m)}_k:=\psi_{kp+m-1}^*,
m=1,\dots p$ and $k\in Z$, where we call $m$ the {\em color} of
fermions $\psi^{(m)}_k,{\psi^*}^{(m)}_k$. Then, expectation value
(\ref{tauhyp1}) factorizes into the product of $p$ expectation
values over all colors.

Let $(\alpha_1,\alpha_2,\dots |\beta_1,\beta_2,\dots)$ be the
Frobenius notation of the partition $\lambda$. We introduce
\begin{equation}\label{r^{(m)}_lambda}
r^{(m)}_\lambda(k)=r_{\tilde \lambda}(k+m-p+1) \ ,\quad m=1,\dots
,p \ ,
\end{equation}
where ${\tilde \lambda}$ is the partition
$(p\alpha_1,p\alpha_2,\dots |p\beta_1,p\beta_2,\dots )$.

Let us consider a set of partitions
$\lambda^{(m)}=(\alpha_1^{(m)},\alpha_2^{(m)},
\dots,\alpha_{k^{(m)}}^{(m)} |\beta_1^{(m)},\beta_2^{(m)},\dots
,\beta_{k^{(m)}}^{(m)}), m=1,\dots,p$. We consider the numbers
$\{\alpha_{k_i}^{(m)}p+m-1 |i=1,\dots,k^{(m)}, m=1,\dots,p\}$, and
label them in such a way that $\alpha_1>\alpha_2>\cdots
>\alpha_K>0$, where $K=k^{(1)}+\cdots +k^{(p)} $ (by the construction,
the decreasing order is strict ). Similarly, we consider the
numbers $\{\beta_{k_i}^{(m)}p+m-p |i=1,\dots,k^{(m)},
m=1,\dots,p\}$, and label them in the same way:
$\beta_1>\beta_2>\cdots
>\beta_K>0$. Let us
consider the partition $\lambda=(\alpha_1,\dots,\alpha_K |\beta_1,
\dots,\beta_K)$. Similarly to (\ref{tauhyp1tN}), we obtain
\begin{equation}\label{schurtildet}
\prod_{m=1}^p s_{\lambda^{(m)}} ({\bf { t}})=(-)^{c_\lambda}
s_\lambda ({\bf t}^{[p]}) ,\quad s_\lambda
(0,\dots,0,1,0,\dots,0,0,\dots)=(-)^{c_\lambda}\prod_{m=1}^p
H^{-1}_{\lambda^{(m)}} \ ,
\end{equation}
where the sign factor $(-)^{c_\lambda}$ originates from the
reordering of the fermions in the expectation, and can be obtained
as $ sign \prod_{i,j}(h_i^{(m_2)}-h_j^{(m_1)})$, here we keep the
notation $h_i$ of for the so-called shifted weights
$\ell(\lambda)+\lambda_i-i$. We write each integer
$h$ as a product of $p$ groups of $\frac{\ell(\lambda)}{p}$
integers $h^{(m)}$ with $m=0,\dots ,p-1$ denoting their congruence
modulo $p$.

\section{Appendix C. Applications of the tau functions of the
hypergeometric type}

Here, we only point out the correspondence between the tau
functions of hypergeometric type and objects introduced in
different papers.

(A)  Two-dimensional Yang-Mills theory \cite{Mig} and a  particular case of the so-called non-local Yang-Mills theory \cite{SK} (which are both considered on a disk) result in the following partition function:
\begin{equation}\label{2DYM}
Z(U_1,U_2)=\sum_{R}\xi_R(U_1^{-1})\xi_R(U_2^{-1}) e^{A\sum_{m=1}^\infty a_mC_m(R)}
\end{equation}
where $a_m$ are some constants (in 2D Yang-Mills theory
all of $a_m$ vanish except $a_2$ which is related to the coupling constant \cite{Mig}), and where the summation runs over irreducible representations of the gauge group;
$U_1$ and $U_2$ are the path-ordered exponentials (Wilson loops) of the gauge field on the
boundaries,  $A$ is the area of
the disk, and $C_m,\ m=1,2,\dots$, are the Casimirs
of the gauge group $SU(N)$:
\begin{equation}\label{higherCasimirs}
{C}_m(R)=\sum_{i=1}^N\left[(\lambda_i+N-i)^m-(N-i)^m\right]
\end{equation}

Chosing
\begin{equation}\label{Ut}
mt_m=Tr U_1^m,\quad mt_m^*=Tr U_2^m
\end{equation}
we find that (\ref{2DYM}) is of form (\ref{htf}).

(B) The generating function for double Hurwitz numbers
  was introduced in \cite{O} in the study of the Gromov-Witten
potential, and has form (\ref{htf})
\begin{equation}\label{Hurwitz}
\tau(P,P',\beta,q)=\sum_\lambda q^{|\lambda|}e^{\beta f_2(\lambda)}s_\lambda(P)s_\lambda(P')
\end{equation}
where $P$ and $P'$ are related to ${\bf t}$ and ${\bf t}^*$,
and where $f_2$ is
\begin{equation}\label{fokounkov}
f_2(\lambda)=
\frac 12 \sum_i\left[(\lambda_i-i+\frac 12)^2 -
(-i+\frac 12)^2 \right]
\end{equation}

Choosing ${\bf t}={\bf t}^*=e^{\frac u2} {\bf t}_\infty
:=(e^{\frac u2},0,0,\dots )$ (as was done in \cite{O},\cite{OP}),
one obtains  the following version of 1D Toda lattice  from
(\ref{rToda'}):
\begin{equation}\label{rTodaeq}
e^{-u}\frac{d^2}{du^2} \phi_n= r(n)e^{\phi_{n-1}-\phi_{n}}-
r(n+1)e^{\phi_{n}-\phi_{n+1}}
\end{equation}
Under a special reduction: $n=\log t_1t_1^*+const$ (which reminds
Benjamin-Ono reduction \cite{LOPZ}) one obtains so-called
equivariant Toda equation, see \cite{O}.
In this case
the $\beta$ is the ${\tilde t}_2$ of (\ref{ThA}), while all
other ${\tilde t}_k$ vanish.

(C) Supersymmetric gauge theory \cite{N},\cite{LMN}. In \cite{N}
the softly broken supersymmetric $\textmd{N}=4$ theory, which is
supersymmetric  $\textmd{N}=2$ gauge theory (with the gauge group
$U(N)$) is considered. The evaluation of the $4D$ Euclidean path
integral in the $\Omega$ background yields in the $U(1)$ case (the
Abelian gauge theory) the partition function
\begin{equation}
Z(\Lambda)=e^{-\frac{1}{12}}\sum_{\bf k}\Lambda^{2|{\bf
k}|}\mu({\bf k})^2,\quad \mu({\bf
k})=\prod_{i<j}\left(\frac{k_i-k_j+j-i}{j-i}\right)
\end{equation}
It is related to our notations as follows: ${\bf
k}=(k_1,k_2,\dots)\to\lambda=(\lambda_1,\lambda_2,\dots)$,
$\Lambda^2=r$. Then, it is the tau function of hypergeometric
type, where ${\bf t}={\bf t}^*={\bf t}_\infty$. The case $N>1$ is
also related to the tau function (\ref{htf}) after certain
transformation. In \cite{N} the 5D supersemmetric theory with the
eight supercharges, compactified on the circle $S^1$ of
circumference $\beta$ was also studied. In $U(1)$ case, the
evaluation of the path integral yields the partition function
\begin{equation}
Z(\beta,\Lambda)=\Lambda^{\frac{1-N^2}{12}}
e^{\gamma_\hbar(0|\beta,\Lambda)}\sum_{\bf
k}(\beta\Lambda)^{2|{\bf k}|}\mu({\bf k})^2,\quad \mu({\bf
k})=\prod_{i<j}\left(\frac{\sinh\frac{\beta \hbar}{2}
(k_i-k_j+j-i)}{\sinh\frac{\beta \hbar}{2} (j-i)}\right)
\end{equation}
It is tau function (\ref{vactau}) with ${\bf t}={\bf
t}^*={\bf t}(\infty,q)$.

(D) The intersection numbers on Hilbert schemes \cite{LQW}. The
identification is the following. The notations $m,t,s$ and $x$
corresponds to the respective notations $n,{\bf t},{\bf t}^*$ and
$T$ in the present paper. The notation $\tau(t,s,x,m)$ in
\cite{LQW} corresponds to the $\tau(n,{\bf t},T,{\bf t}^*)$. The
$\tau(t,s,x,m)$ in the paper \cite{LQW} is the generating function
of the equivariant intersection numbers on Hilbert schemes of
points on the affine plane.

(E) An example of solvable matrix integral \cite{1'}
 ($\sigma$ is the
partition $ (m,\dots , m), \ell(\sigma)=n$)
\begin{equation}\label{Wetting}
\int_{U(n)}\det U^{\mp m}\det(1- X U)^{-a}\det(1- U^{-1}Y)^{-b}
d_*U= \frac{1} {(n)_\sigma}\left.\left\{({ a})_\sigma\det X^{m}
\atop ({\tilde a})_\sigma\det Y^{m}\right.\right\}
{}_2F_1\left.\left({\tilde a},{\tilde b} \atop n+m\right |
XY\right),
\end{equation}
where $m\ge 0$ and where ${\tilde a}=a+m,\ {\tilde b}=b$ for the
upper sign and where ${\tilde a}=a,\ {\tilde b}=b+m$ for lower
sign in the l.h.s. of (\ref{Wetting}).  The notations
$(n)_\sigma,(a)_\sigma$ one finds in (\ref{Poch}) . The Gauss
hypergeometric function ${}_2F_1$ of the matrix argument one finds
in (\ref{hZ1z}), it is an example of the hypergeometric tau
function, see \cite{nl64}. In the case $m=0,\ -a=-b=p\in Z_{\ge
0}$ the integral (\ref{Wetting}) was evaluated as  determinant of
$p$ by $p$ matrix in the recent paper \cite{FS} via replica
method. Let us also mark that ${}_2F_1(a,b;c|X)$ with integer
$a,b,c$ solves Painleve V equation \cite{AvM}.

\section{Appendix D. Hypergeometric functions of matrix argument}

The hypergeometric functions, considered below, are examples of
the hypergeometric tau functions \cite{nl64},\cite{pd22}.

{\bf Ordinary hypergeometric functions}. First, let us recall that
the generalized hypergeometric function of one variable $x$ is
defined as
\begin{equation}\label{ordinary}
{}_pF_s\left(a_1,\dots ,a_p; b_1,\dots ,b_s; x \right)
=\sum_{n=0}^\infty\frac {(a_1)_n \cdots (a_p)_n}
             {(b_1)_n \cdots (b_s)_n}
\frac{x^n}{n!}
\end{equation}
where $(a)_n$ is the Pochhammer's symbol
\begin{equation}
(a)_n=\frac {\Gamma(a+n)}{\Gamma(a)}=a(a+1)\cdots (a+n-1) \ .
\end{equation}
This hypergeometric function solves the equation
\begin{equation}\label{ordhypeq}
\left(\prod_{k=0}^s\left(x\frac{d}{dx}+b_k-1\right)-x\prod_{j=1}^p
\left(x\frac{d}{dx}+a_j \right)\right){}_pF_s\left(a_1,\dots ,a_p;
b_1,\dots ,b_s; x \right)=0 \ ,\quad b_0=1
\end{equation}

Given number $q$, $|q|<1$, the so-called basic hypergeometric
series of one variable is defined as
\begin{equation}\label{qordinary}
{}_p\Phi_s\left(a_1,\dots ,a_p; b_1,\dots ,b_s; q, x \right)
=\sum_{n=0}^\infty \frac {(q^{a_1};q)_n \cdots (q^{a_p};q)_n}
             {(q^{b_1};q)_n \cdots (q^{b_s};q)_n}
\frac{x^n}{(q;q)_n} \ .
\end{equation}
Here, $(q^a,q)_n$ is the $q$-deformed  Pochhammer's symbol,
\begin{equation}\label{qPochh}
(b;q)_0=1,\quad (b;q)_n=(1-b)(1-bq^{1})\cdots(1-bq^{n-1}) \ .
\end{equation}
Both series converge for all $x$ in case $p<s+1$. In case $p=s+1$,
they converge for $|x|<1$. We refer these well-known
hypergeometric functions as ordinary hypergeometric functions.

{\bf The multiple hypergeometric series related to Schur
polynomials \cite{V},\cite{KV},\cite{Milne}.}

There are several well-known different multi-variable
generalizations of the hypergeometric series of one variable
\cite{V, KV}. If one replaces the sum over $n$ to the sum over
partitions
 ${\lambda}=(\lambda_1,\lambda_2,\dots)$, and replaces the single
variable $x$ to a Hermitian matrix $X$, he obtains the
generalization of the hypergeometric series, which is called the
{\em hypergeometric function of matrix argument ${\bf X}$ with
indices $\bf{a}$ and $\bf{b}$} \cite{GR},\cite{V}:
\begin{eqnarray}\label{hZ1z}
{}_pF_s\left.\left(a_1,\dots ,a_p\atop b_1,\dots
,b_s\right|{{\bf{X}}} \right) =\sum_{{\lambda}\atop
l({\lambda})\le N} \frac {(a_1)_{{\lambda}} \cdots
(a_p)_{{\lambda}}}
                     {(b_1)_{{\lambda}} \cdots (b_s)_{{\lambda}}}
\frac{Z_{\lambda}({\bf X})} {|{\lambda}|!} \ .
\end{eqnarray}
Here, the sum ranges all different partitions
${\lambda}=\left(\lambda_1,\lambda_2,\dots,\lambda_k\right)$,
where $\lambda_1\ge \lambda_2\ge \cdots \ge \lambda_k$,  whose
length $\ell({\lambda})=k\le N$.  The ${\bf X}$ is a Hermitian
$N\times N$ matrix, and $Z_{\lambda}({\bf X})$ is the zonal
spherical polynomial for the symmetric spaces of the following
types: $GL(N,R)/SO(N)$, $GL(N,C)/U(N)$, or $GL(N,H)/Sp(N)$,
 see \cite{KV}. The definition of symbol $(a)_{\lambda}$ depends
on the choice of the symmetric space:
\begin{equation}
(a)_{\lambda}=(a)_{\lambda_1}(a-\frac{1}{\alpha})_{\lambda_2}
\cdots(a-\frac{k-1)}{\alpha})_{\lambda_k} \ ,\quad (a)_{\bf 0}=1 \
,
\end{equation}
where we respectively substitute the values of the parameter
$\alpha$ as $\alpha=2$, $\alpha=1$ and $\alpha =\frac12$ for the
symmetric spaces $GL(N,R)/SO(N)$, $GL(N,C)/U(N)$, and
$GL(N,H)/Sp(N)$. Function (\ref{hZ1z}) actually depends on
eigenvalues of matrix $X$, which are
 ${\bf x}^N=(x_1,\dots,x_N)$ .
We consider only the case of the symmetric space $GL(N,C)/U(N)$.
Then, zonal spherical polynomial $Z_{\lambda}({\bf X})$ is
proportional to the Schur function $s_{\lambda}(x_1,x_2,...,x_N)$
corresponding to the partition ${\lambda}$ \cite{Mac}.

For this choice of the symmetric space, the hypergeometric
function can be rewritten as follows:
\begin{eqnarray}\label{hZ1}
{}_pF_s\left.\left(a_1,\dots ,a_p\atop b_1,\dots
,b_s\right|{{\bf{X}}} \right) = \sum_{{\lambda}\atop
l({\lambda})\le N} \frac {(a_1)_{{\lambda}} \cdots
(a_p)_{{\lambda}}}
                     {(b_1)_{{\lambda}} \cdots (b_s)_{{\lambda}}}
\frac{s_{\lambda}({\bf x}^N)} {H_{\lambda}} \ ,
\end{eqnarray}
where $H_{\lambda}$ is the-product-of-hook's-length:
\begin{equation}\label{hookprod}
H_{{\lambda}}=\prod_{(i,j)\in {\lambda}} h_{ij} \ , \quad
h_{ij}=(\lambda_i+\lambda'_j-i-j+1)
\end{equation}
and
\begin{equation}\label{Poch}
(a)_{\lambda}=(a)_{\lambda_1}(a-1)_{\lambda_2}
\cdots(a-k+1)_{\lambda_k} \ , \quad (a)_{\bf 0}=1
\end{equation}
Taking $N=1$, we get (\ref{ordinary}).

It was shown \cite{nl64},\cite{pd22}, that  hypergeometric
function (\ref{hZ1}) is  tau function (\ref{exex'}):
\begin{equation}\label{hZ1tau}
{}_pF_s\left.\left(a_1+n,\dots ,a_p+n\atop b_1+n,\dots
,b_s+n\right|{{\bf{X}}} \right) =\tau_r(n,{\bf t},{\bf t}_\infty)
\ ,
\end{equation}
where
\begin{equation}\label{hZ1r}
r(k)= {\prod_{i=1}^p(k+a_i)}{\prod_{i=1}^s(k+b_i)^{-1}}
\end{equation}
and
\begin{equation}\label{hZ1tX}
t_m= \textrm{Tr}\ {\bf X}^m \ , \quad m=1,2,3,\dots
\end{equation}
Thus, the hypergeometric function (\ref{hZ1}) is a subject of
(\ref{ThA}) of the Theorem 1.

Let $|q|<1$. The multiple {\em basic} hypergeometric series
related to the Schur functions were suggested by L.G.Macdonald and
studied by S.Milne \cite{Milne}. These series are
\begin{eqnarray}\label{vh}
{}_p\Phi_s\left(a_1,\dots ,a_p;b_1,\dots ,b_s;q,{\bf x}^N\right)
=\sum_{{\lambda}\atop l({\lambda})\le N}\frac
{(q^{a_1};q)_{{\lambda}} \cdots (q^{a_p};q)_{{\lambda}}}
{(q^{b_1};q)_{{\lambda}}\cdots (q^{b_s};q)_{{\lambda}}}
\frac{q^{n({\lambda})}}{H_{{\lambda}}(q)} s_{{\lambda}} \left(
{\bf x}^N \right) \ ,
\end{eqnarray}
where the sum ranges all different partitions
${\lambda}=\left(\lambda_1,\lambda_2,\dots,\lambda_k\right)$,
where $\lambda_1\ge \lambda_2\ge \cdots \ge \lambda_k \ge 0$,
$k\le |{\lambda}|$, and whose length $l({\lambda})=k\le N$. The
multiplier $(q^c;q)_{\lambda}$, associated with a partition
${\lambda}$,
 is expressed in terms of the $q$-deformed Pochhammer's Symbols
$(q^c;q)_{n}$ (\ref{qPochh}):
\begin{equation}
(q^c;q)_{\lambda}=(q^c
;q)_{\lambda_1}(q^{c-1};q)_{\lambda_2}\cdots
(q^{c-k+1};q)_{\lambda_k} \ .
\end{equation}
The multiple $q^{n({\lambda})}$, defined on a partition
${\lambda}$, is
\begin{equation}\label{n(n)}
q^{n({\lambda})}=q^{\sum_{i=1}^k (i-1)\lambda_i} \ ,
\end{equation}
and $q$-deformed the-product-of-hook's-length (the so-called hook
polynomial) $H_{{\lambda}}(q)$ is
\begin{eqnarray}\label{hp}
H_{{\lambda}}(q)=\prod_{(i,j)\in {\lambda}}
\left(1-q^{h_{ij}}\right) \ , \quad
h_{ij}=(\lambda_i+\lambda'_j-i-j+1) \ ,
\end{eqnarray}
where ${\lambda}'$ is the conjugated partition.

For $N=1$, we get (\ref{qordinary}). \\

Let us note that in the limit $q \to 1$ series (\ref{vh}) (up to a
multiplier)
reduces to (\ref{hZ1}), see \cite{KV}.\\

It was shown \cite{nl64},\cite{pd22}, that the hypergeometric
function (\ref{hZ1}) is tau function (\ref{exex'}):
\begin{equation}\label{Phitau}
{}_p\Phi_s\left(a_1,\dots ,a_p;b_1,\dots ,b_s;q,{\bf X}\right)
=\tau_r(n,{\bf t},{\bf t}(\infty,q))
\end{equation}
where ${\bf t}(\infty,q)$ is given by (\ref{choicetinftyq}), and
where
\begin{equation}\label{hZ1tX'}
t_m= \textrm{Tr}\ {\bf X}^m \ , \quad m=1,2,3,\dots \ ,
\end{equation}
\begin{equation}\label{rq}
r(n)= \frac{\prod_{i=1}^p (1-q^{a_i+n})}
{\prod_{i=1}^s(1-q^{b_i+n})}
\end{equation}
Thus, the Milne hypergeometric function (\ref{vh}) is also a
subject of (\ref{ThC}) of the Theorem 1.

\section{Acknowledgements}

The author thanks Anton Zabrodin, Andrei Semenovich Losev, Sergei
Natanzon, Igor Loutsenko and, most of all, Nikita Nekrasov  for
helpful discussions. He thanks M.Boiti, T.Degasperis, P.Santini,
G.Helmink and J. van de Leur for the discussions and the
organization of the lecture in Galipolli and the seminars at Rome,
Utrecht and Twenty universities, where the present paper was
started. The work was supported by the Russian Foundation for
Fundamental Researches (Grant No 01-01-00548),
and the Program of Russian Academy of Science "Mathematical Methods in Nonlinear Dynamics".

\end{document}